\documentclass{aastex}%
\usepackage{amsmath}
\usepackage{amssymb}%
\usepackage{amsfonts}%
\usepackage{graphicx}

\begin{document}
{\LARGE Streaming cold cosmic ray back-reaction and thermal instabilities
across the background magnetic field}

\bigskip

\ \ \ \ \ \ \ \ \ \ \ \ \ \ \ \ \ \ \ \ \ \ {\large Anatoly K. Nekrasov}$^{1}
${\large \ and Mohsen Shadmehri}$^{2}$

$^{1}$ Institute of Physics of the Earth, Russian Academy of Sciences, 123995
Moscow, Russia;

anekrasov@ifz.ru, nekrasov.anatoly@gmail.com

$^{2}$ Department of Physics, Golestan University, Basij Square, Gorgan, Iran;

m.shadmehri@gu.ac.ir

\bigskip

\ \ \ \ \ \ \ \ \ \ \ \ \ \ \ \ \ \ \ \ \ \ \ \ \ \ \ \ \ \ \ \ \ \ \ \ \ \

\ \ \ \ \ \ \ \ \ \ \ \ \ \ \ \ \ \ \ \ \ \ \ \ \ \ \ \ \ \ \ \ \ \ \ \ \ \ \ \ \ \ \ \ \ \ \ \ \ \ ABSTRACT

Using the multi-fluid approach, we investigate streaming and thermal
instabilities of the electron-ion plasma with homogeneous cold cosmic rays
drifting perpendicular to the background magnetic field. Perturbations across
the magnetic field are considered. The back-reaction of cosmic rays resulting
in the streaming instability is taken into account. The thermal instability is
shown not to be subject to the action of cosmic rays in the model under
consideration. The dispersion relation for the thermal instability has been
derived which includes sound velocities of plasma and cosmic rays, Alfv\'{e}n
and cosmic ray drift velocities. The relation between these parameters
determines the kind of thermal instability from Parker's to Field's type
instability. The results obtained can be useful for a more detailed the
investigation of electron-ion\textbf{\ }astrophysical objects\textbf{\ }such
as galaxy clusters including the dynamics of streaming cosmic rays.

\bigskip


\bigskip

\section{INTRODUCTION}

The cosmic rays are an important ingredient in some of the astrophysical
environments (e.g. Zweibel 2003). They are capable of affecting the dynamics
of the astrophysical plasma media leading to plasma heating, increasing the
level of ionization, driving outflows, modifying shocks, and so on (e.g.
Field, Goldsmith \& Habing 1969; Zweibel 2003; Guo \& Oh 2008; Everett et al.
2008; Beresnyak, Jones \& Lazarian 2009; Samui, Subramanian \& Srianand 2010;
En\ss lin et al. 2011). Ionization by cosmic rays has a vital role in star
formation near the Galactic center (e.g. Yusef-Zadeh, Wardle \& Roy 2007) and
in the dead zone of protoplanetary disks (Gammie 1996).

The thermal instability (Field 1965) has been used to explain existence of the
cold dense structures in the interstellar (e.g. Field 1965; Begelman \& McKee
1990; Koyama \& Inutsuka 2000; Hennebelle \& P\'{e}rault 2000;
S\'{a}nchez-Salcedo, V\'{a}zquez-Semadeni \& Gazol 2002; V\'{a}zquez-Semadeni
et al. 2006; Fukue \& Kamaya 2007; Inoue \& Inutsuka 2008; Shadmehri,
Nejad-Asghar \& Khesali 2010) and intracluster (ICM; e.g. Field 1965; Mathews
\& Bregman 1978; Balbus \& Soker 1989; Loewenstein 1990; Bogdanovi\'{c} et al.
2009; Parrish, Quataert \& Sharma 2009; Sharma, Parrish \& Quataert 2010)
media. For example, molecular filaments \ are seen in galaxy clusters with
short ($\lesssim$1 Gyr) cooling times (e.g. Conselice, Gallagher \& Wyse 2001;
Salom\'{e} et al. 2006; Cavagnolo et al. 2008; O'Dea et al. 2008).

In galaxy clusters, cosmic rays are wide spread (e.g. Guo \& Oh 2008;
En\ss lin et al. 2011). Therefore, they could exert influence on the thermal
instability. In particular, including cosmic rays is required to explain the
atomic and molecular lines observed in filaments in clusters of galaxies
(Ferland et al. 2009). Such an investigation has been performed by Sharma,
Parrish \& Quataert (2010) in the framework of the magnetohydrodynamic (MHD)
equations where cosmic rays have been considered as a second fluid having the
velocity of the thermal plasma. Numerical analysis has shown that the cosmic
ray pressure can play an important role in the dynamics of cold filaments
making them much more elongated along the magnetic field lines than the Field length.

However, the relativistic cosmic rays can have a drift velocity of the order
of the speed of light and temperature larger than the particle rest energy.
The interaction of such particles with the thermal plasma can not be
considered in the framework of the conventional MHD. The cosmic ray drift
current results in arising of the return current provided by the background
plasma (e.g. Achterberg 1983; Bell 2004, 2005; Riquelme \& Spitkovsky 2009,
2010). The possible role of this effect in the generation of thermal
instability needs to be considered. There is also another important issue like
the amplification of magnetic fields. The classical cyclotron resonant
instability has been proposed long time ago to explain this process (Kulsrud
\& Pearce 1969). However, this mechanism has turned out to be unable to
provide sufficient energy in the shock upstream plasma. In order to resolve
this problem, just recently a new non-resonant instability has been proposed
that may provide a much higher energy (Bell 2004; see also Zweibel 2003). This
instability known as the Bell instability has been confirmed by non-linear
numerical simulations (Riquelme \& Spitkovsky 2009). Subsequent works extended
this instability into various directions by considering partially ionized
media (Reville et al. 2007) and thermal plasma effects (Zweibel \& Everett
2010). However, the works sited above, except for the paper by Bell (2005),
have been restricted to the cosmic ray drift velocity and perturbations
parallel to the initial magnetic field. In his paper, Bell (2005) has derived
the general dispersion relation for arbitrary orientation of the background
magnetic field, cosmic ray current, and direction of perturbations. The
dispersion relation obtained describes instability due to the return plasma
current. In the paper by Riquelme \& Spitkovsky (2010), the case in which the
cosmic ray current is perpendicular to the initial magnetic field has also
been considered. In this case, cosmic rays can be magnetized in a way that
their Larmor radius defined by the longitudinal thermal velocity (Zweibel
2003; Bell 2004) is smaller than the typical length scales of the system.
Riquelme \& Spitkovsky (2010) have studied this perpendicular current-driven
instability analytically in the linear regime and numerically. Their growth
rate was similar to that of the cosmic ray current-driven instability by Bell
(2004). But these authors have not included the cosmic ray back-reaction analytically.

The thermal instability in galaxy clusters in the multi-fluid approach has
been considered by Nekrasov (2011, 2012). Effects related to cosmic rays were
not included in these papers. Although the original Bell instability was
proposed to explain magnetic field amplification in a shock, just recently
Nekrasov and Shadmehri (2012) extended the instability to a multi-fluid case
in which the thermal effects are also considered along with the presence of
streaming cold cosmic rays.\textbf{\ }A geometry was considered in which
homogeneous cosmic rays drift across the background magnetic field and
perturbations arise along the latter. Such a geometry was analogous to that
treated by Riquelme \& Spitkovsky (2010). The cosmic ray back-reaction has
been included and the growth rate has been obtained which is much larger than
that for the Bell instability and perpendicular current-driven instability by
Riquelme \& Spitkovsky (2010). These\textbf{\ }findings motivated us to
investigate the case in which perturbations arise transversely to the ambient
magnetic field in the directions both along and across the cosmic ray current.
As it is followed from the paper by Bell (2005) using the MHD equations, a
streaming instability does not exist for such a geometry. However, this result
is incorrect in the multi-fluid consideration that is shown in this paper and
has been obtained earlier (see for example Nekrasov (2007)). We include the
induced return current of the background plasma and back-reaction of cosmic
rays. With such an approach, the dispersion relations are derived and the
growth rates are found analytically. We also consider possible effects of
cosmic rays on the thermal instability. For simplicity, we ignore the action
of gravity as it has been done by Sharma, Parrish \& Quataert (2010). The
effects of the gravitational field have been investigated in detail by the
multi-fluid approach in papers by Nekrasov \& Shadmehri (2010, 2011). Thus,
our present study extends previous analytical studies by considering not only
the thermal effects but the currents driven by cosmic rays and their back-reaction.

The paper is organized as follows. Section 2 contains the fundamental
equations for plasma, cosmic rays, and electromagnetic fields used in this
paper. Equilibrium state is discussed in Section 3. Wave equations are given
in Section 4. In Sections 5 and 6, the dispersion relations including the
plasma return current, cosmic ray back-reaction, and the terms describing the
thermal instability are derived and their solutions are found for
perturbations along and across the cosmic ray current, respectively.
Discussion of important results obtained and possible astrophysical
implications are provided in Section 7. Conclusive remarks are summarized in
Section 8.

\bigskip

\section{BASIC\ EQUATIONS FOR\ PLASMA\ AND\ COSMIC\ RAYS}

The fundamental equations for a plasma that we consider here are the
following:%
\begin{equation}
\frac{\partial\mathbf{v}_{j}}{\partial t}+\mathbf{v}_{j}\cdot\mathbf{\nabla
v}_{j}=-\frac{\mathbf{\nabla}p_{j}}{m_{j}n_{j}}+\frac{q_{j}}{m_{j}}%
\mathbf{E+}\frac{q_{j}}{m_{j}c}\mathbf{v}_{j}\times\mathbf{B},
\end{equation}
the equation of motion,%
\begin{equation}
\frac{\partial n_{j}}{\partial t}+\mathbf{\nabla}\cdot n_{j}\mathbf{v}_{j}=0,
\end{equation}
the continuity equation,%
\begin{equation}
\frac{\partial T_{i}}{\partial t}+\mathbf{v}_{i}\cdot\mathbf{\nabla}%
T_{i}+\left(  \gamma-1\right)  T_{i}\mathbf{\nabla}\cdot\mathbf{v}%
_{i}=-\left(  \gamma-1\right)  \frac{1}{n_{i}}%
\mathcal{L}%
_{i}\left(  n_{i},T_{i}\right)  +\nu_{ie}^{\varepsilon}\left(  n_{e}%
,T_{e}\right)  \left(  T_{e}-T_{i}\right)
\end{equation}
and%
\begin{equation}
\frac{\partial T_{e}}{\partial t}+\mathbf{v}_{e}\cdot\mathbf{\nabla}%
T_{e}+\left(  \gamma-1\right)  T_{e}\mathbf{\nabla}\cdot\mathbf{v}%
_{e}=-\left(  \gamma-1\right)  \frac{1}{n_{e}}%
\mathcal{L}%
_{e}\left(  n_{e},T_{e}\right)  -\nu_{ei}^{\varepsilon}\left(  n_{i}%
,T_{e}\right)  \left(  T_{e}-T_{i}\right)
\end{equation}
are the temperature equations for ions and electrons. In Equations (1) and
(2), the index $j=i,e$ denotes the ions and electrons, respectively. Notations
in Equations (1)-(4) are the following: $q_{j}$ and $m_{j}$ are the charge and
mass of species $j$, $\mathbf{v}_{j}$ is the hydrodynamic velocity, $n_{j}$ is
the number density, $p_{j}=n_{j}T_{j}$ is the thermal pressure, $T_{j}$ is the
temperature, $\nu_{ie}^{\varepsilon}(n_{e},T_{e})$ ($\nu_{ei}^{\varepsilon
}\left(  n_{i},T_{e}\right)  $) is the frequency of the thermal energy
exchange between ions (electrons) and electrons (ions) being $\nu
_{ie}^{\varepsilon}(n_{e},T_{e})=2\nu_{ie}$, where $\nu_{ie}$ is the collision
frequency of ions with electrons (Braginskii 1965), $n_{i}\nu_{ie}%
^{\varepsilon}\left(  n_{e},T_{e}\right)  =n_{e}\nu_{ei}^{\varepsilon}\left(
n_{i},T_{e}\right)  $, $\gamma$ is the ratio of the specific heats,
$\mathbf{E}$\textbf{\ }and $\mathbf{B}$ are the electric and magnetic fields,
and $c$ is the speed of light in vacuum. For simplicity, here we do not take
into account collisions between the ions and electrons in the momentum
equation. This effect for the thermal instability has been treated by Nekrasov
(2011, 2012), where, in particular, conditions, under which such collisions
can be neglected, have been found. However, the thermal exchange should be
included because it must be compared with the dynamical time. The cooling and
heating of plasma species in Equations (3) and (4) are described by the
function $%
\mathcal{L}%
_{j}(n_{j},T_{j})=n_{j}^{2}\Lambda_{j}\left(  T_{j}\right)  -n_{j}\Gamma_{j}$,
where $\Lambda_{j}$ and $\Gamma_{j}$ are the cooling and heating functions,
respectively. The form of this function has a certain deviation from the
usually used cooling-heating function $\pounds $ (Field 1965). Both functions
are connected to each other via the equality $%
\mathcal{L}%
_{j}\left(  n_{j},T_{j}\right)  =m_{j}n_{j}\pounds _{j}$. Our choice is
analogous to those of Begelman \& Zweibel (1994), Bogdanovi\'{c} et al.
(2009), Parrish, Quataert \& Sharma (2009). The function $\Lambda_{j}\left(
T_{j}\right)  $ can be found, for example, in Tozzi \& Norman (2001). We do
not take into account the transverse thermal fluxes in the temperature
equations, which are small in the weekly collisional plasma (Braginskii 1965)
being considered in this paper.

Equations for relativistic cosmic rays which can be in general both protons
and electrons we use in the form (e.g. Lontano, Bulanov \& Koga 2002)
\begin{equation}
\frac{\partial\left(  R_{cr}\mathbf{p}_{cr}\right)  }{\partial t}%
+\mathbf{v}_{cr}\cdot\mathbf{\nabla}\left(  R_{cr}\mathbf{p}_{cr}\right)
=-\frac{\mathbf{\nabla}p_{cr}}{n_{cr}}+q_{cr}\left(  \mathbf{E+}\frac{1}%
{c}\mathbf{v}_{cr}\times\mathbf{B}\right)  ,
\end{equation}%
\begin{equation}
\left(  \frac{\partial}{\partial t}+\mathbf{v}_{cr}\cdot\mathbf{\nabla
}\right)  \left(  \frac{p_{cr}\gamma_{cr}^{\Gamma_{cr}}}{n_{cr}^{\Gamma_{cr}}%
}\right)  =0,
\end{equation}
where%
\begin{equation}
R_{cr}=1+\frac{\Gamma_{cr}}{\Gamma_{cr}-1}\frac{T_{cr}}{m_{cr}c^{2}}.
\end{equation}
In these equations, $\mathbf{p}_{cr}=\gamma_{cr}m_{cr}\mathbf{v}_{cr}$ is the
momentum of a cosmic ray particle having the rest mass $m_{cr}$ and velocity
$\mathbf{v}_{cr}$, $q_{cr}$ is its charge, $p_{cr}=\gamma_{cr}^{-1}%
n_{cr}T_{cr}$ is the kinetic pressure, $n_{cr}$ is the number density in the
laboratory frame, $\Gamma_{cr}$ is the adiabatic index, $\gamma_{cr}=\left(
1-\mathbf{v}_{cr}^{2}/c^{2}\right)  ^{-1/2}$ is the relativistic factor. The
continuity equation is the same as Equation (2) for $j=cr$. Equation (7) can
be used for both cold nonrelativistic, $T_{cr}\ll$ $m_{cr}c^{2}$, and hot
relativistic, $T_{cr}\gg$ $m_{cr}c^{2}$, cosmic rays. In the first (second)
case, we have $\Gamma_{cr}=5/3$ ($4/3$) (Lontano, Bulanov \& Koga 2002). The
general form of the value $R_{cr}$, which is valid for any relations between
$T_{cr}$ and $m_{cr}c^{2}$, can be found e.g. in Toepfer (1971) and
Dzhavakhishvili and Tsintsadze (1973).

Equations (1)-(4), (5), and (6) are solved together with Maxwell's equations
\begin{equation}
\mathbf{\nabla\times E=-}\frac{1}{c}\frac{\partial\mathbf{B}}{\partial t}%
\end{equation}
and
\begin{equation}
\mathbf{\nabla\times B=}\frac{4\pi}{c}\mathbf{j+}\frac{1}{c}\frac
{\partial\mathbf{E}}{\partial t},
\end{equation}
where $\mathbf{j=j}_{pl}+\mathbf{j}_{cr}=\sum_{j}q_{j}n_{j}\mathbf{v}%
_{j}+\mathbf{j}_{cr}$. Below, we first consider an equilibrium state in which
there is a stationary cosmic ray current.

\bigskip

\section{EQUILIBRIUM STATE}

We will consider a uniform plasma embedded in the uniform magnetic field
$\mathbf{B}_{0}$ (subject $0$ here and below denotes background parameters)
directed along the $z$-axis. We assume that the plasma in equilibrium is
penetrated by a uniform beam of cosmic rays having the uniform streaming
velocity $\mathbf{u}_{cr}$ along the $y$-axis. It is reasonable to suppose
that in such uniform model the magnetic field due to cosmic rays will be
absent. This picture is analogous to the consideration of the gravitational
potential in the equilibrium state in an infinite uniform medium having a
constant mass density. Then we obtain from Equation (9)
\begin{equation}
\sum_{j}q_{j}n_{j0}\mathbf{v}_{j0}+\mathbf{j}_{cr0}\mathbf{+}\frac{1}{4\pi
}\frac{\partial\mathbf{E}_{0}}{\partial t}=0.
\end{equation}
From Equation (1), we easily find in the equilibrium state%
\begin{equation}
\mathbf{v}_{e0}=\mathbf{v}_{0}=c\frac{\mathbf{E}_{0}\times\mathbf{B}_{0}%
}{B_{0}^{2}},\mathbf{v}_{i0}=\mathbf{v}_{0}+\frac{c}{\omega_{ci}B_{0}}%
\frac{\partial\mathbf{E}_{0}}{\partial t},v_{e0z}=v_{i0z}=0,
\end{equation}
where we have assumed that $\partial/\partial t\ll\omega_{cj}$, $\omega_{cj}$
$=q_{j}B_{0}/m_{j}c$ is the cyclotron frequency. Analogously, we obtain from
Equation (5) under condition $R_{cr}\gamma_{cr}\partial/\partial t\ll
\omega_{cr}$ ($\omega_{cr}=q_{cr}B_{0}/m_{cr}c$)
\begin{equation}
\mathbf{v}_{cr0}=\mathbf{v}_{0}+\mathbf{u}_{cr}.
\end{equation}
In Equation (12), we have neglected the polarization drift of cosmic rays (the
second term on the right-hand side for the ion velocity in Equation (11)).
This can be done for the approximate condition $n_{i0}\gg n_{cr0}$, if the
cosmic rays are protons (see below), which is always satisfied. Using
Equations (11) and (12), we will find the current $\mathbf{j}_{0}$%
\begin{equation}
\mathbf{j}_{0}=\frac{q_{i}n_{i0}c}{\omega_{ci}B_{0}}\frac{\partial
\mathbf{E}_{0}}{\partial t}+q_{cr}n_{cr0}\mathbf{u}_{cr},
\end{equation}
where we have taken into account the condition of quasi-neutrality
$q_{i}n_{i0}+q_{e}n_{e0}+q_{cr}n_{cr0}=0$ (the number density $n_{cr}$ is the
one in the laboratory frame). This condition is satisfied in astrophysical
plasmas due to cosmic ray charge neutralization from the background
environment (Alfv\'{e}n 1939). Substituting Equation (13) into Equation (10),
we obtain%
\begin{equation}
\frac{\partial\mathbf{E}_{0}}{\partial t}=-4\pi q_{cr}n_{cr0}\mathbf{u}%
_{cr}\frac{c_{Ai}^{2}}{c_{Ai}^{2}+c^{2}},
\end{equation}
where $c_{Ai}=\left(  B_{0}^{2}/4\pi m_{i}n_{i0}\right)  ^{1/2}$ is the ion
Alfv\'{e}n velocity. Usually, the inequality $c^{2}\gg c_{Ai}^{2}$ is
satisfied. In this case, Equation (14) coincides with the corresponding
equation given in the paper by Riquelme \& Spitkovsky (2010). Substituting
Equation (14) into Equation (11) for ions, we find the return plasma current%
\begin{equation}
\mathbf{j}_{ret}=q_{i}n_{i0}\left(  \mathbf{v}_{i0}-\mathbf{v}_{0}\right)
=-q_{cr}n_{cr0}\mathbf{u}_{cr},
\end{equation}
which magnitude is equal the cosmic ray current and has the opposite
direction. From Equation (15), it is followed that the induced plasma current
drift velocity $\mathbf{u}_{pl}=\mathbf{v}_{i0}-\mathbf{v}_{0}$ is equal to
$\mathbf{u}_{pl}=-\left(  q_{cr}n_{cr0}/q_{i}n_{i0}\right)  \mathbf{u}_{cr}$.
Using Equation (14), we see that the polarization cosmic ray drift velocity
$\left(  R_{cr}\gamma_{cr}c/\omega_{cr}B_{0}\right)  \partial\mathbf{E}%
_{0}/\partial t$ can be neglected in comparison with $\mathbf{u}_{cr}$ under
condition $m_{i}n_{i0}\gg R_{cr}\gamma_{cr}m_{cr}n_{cr0}$. If the cosmic rays
(protons) are not too relativistic, this condition is satisfied. Below, the
plasma drift velocity $\mathbf{u}_{pl}$ will be also taken into account as
$\mathbf{u}_{cr}$.

We will consider the case in which background temperatures of the electrons
and ions are equal, i.e. $T_{e0}=T_{i0}=T_{0}$. However, to follow the
symmetric contribution of the ions and electrons in a convenient way, we make
the calculations by assuming different temperatures. In this case, the thermal
equations in equilibrium are given by%

\begin{equation}%
\mathcal{L}%
_{i}\left(  n_{i0},T_{i0}\right)  =%
\mathcal{L}%
_{e}\left(  n_{e0},T_{e0}\right)  =0.
\end{equation}

\bigskip

\section{WAVE\ EQUATIONS}

For perturbations across the background magnetic field, Equations (8) and (9)
give us the following two equations:%
\begin{equation}
c^{2}\left(  \frac{\partial}{\partial t}\right)  ^{-2}\left(  \frac
{\partial^{2}E_{1x}}{\partial y^{2}}-\frac{\partial^{2}E_{1y}}{\partial
x\partial y}\right)  -E_{1x}=4\pi\left(  \frac{\partial}{\partial t}\right)
^{-1}j_{1x}%
\end{equation}
and
\begin{equation}
c^{2}\left(  \frac{\partial}{\partial t}\right)  ^{-2}\left(  -\frac
{\partial^{2}E_{1x}}{\partial x\partial y}+\frac{\partial^{2}E_{1y}}{\partial
x^{2}}\right)  -E_{1y}=4\pi\left(  \frac{\partial}{\partial t}\right)
^{-1}j_{1y},
\end{equation}
where $\mathbf{j}_{1}=\mathbf{j}_{pl1}+\mathbf{j}_{cr1}$ and the subscript $1
$ here and below denotes the perturbed values. The third equation describes
the ordinary electromagnetic wave with $\mathbf{E}_{1}\mathbf{\parallel B}%
_{0}$. The general expressions for the components $j_{pl1x,y}$ and
$j_{cr1x,y}$ are given in the Appendices A and B (Equations (A54)-(A56) and
(B19)-(B21)). These expressions are available for both magnetized and
non-magnetic systems, electron-positron, pair-ion, and dusty plasmas and so
on. In their general form, they are very complicated. Therefore to proceed
analytically, one must apply simplifying assumptions. We are interested in
magnetized systems consisting of electrons, ions, and cosmic rays where
cyclotron frequencies of species are much larger than inverse dynamical times.
In our case, this implies
\begin{align}
\omega_{ci}^{2}  & \gg\left(  \frac{\partial}{\partial t}+u_{pl}\frac
{\partial}{\partial y}\right)  ^{2},\\
\omega_{ccr}^{2}  & \gg\gamma_{cr0}^{4}\left(  \frac{\partial}{\partial
t}+u_{cr}\frac{\partial}{\partial y}\right)  ^{2}\nonumber
\end{align}
(see Equations (A5), (A8), and (B7)). As we have noted above, the cosmic rays
can be both protons and electrons. For ultrarelativistic cosmic rays,
$\gamma_{cr0}\gg1$, the second inequality (19) can be violated. Such a case is
not considered here. Another condition that simplifies the treatment
considerably is to assume the wavelength of perturbations to be much larger
than the Larmor radius of particles $\rho_{j}$%
\begin{align}
1  & \gg\rho_{i}^{2}\mathbf{\nabla}^{2},\\
1  & \gg\rho_{cr}^{2}\gamma_{cr0}\left(  \gamma_{cr0}^{2}\frac{\partial^{2}%
}{\partial x^{2}}+\frac{\partial^{2}}{\partial y^{2}}\right) \nonumber
\end{align}
(see Equations (A41) and (B11)). The additional inequalities for cosmic rays
will be given below. The third simplification is to consider perturbations
along and across the cosmic ray drift velocity separately. The first case is
simpler. Therefore, we begin with its consideration.

\bigskip

\section{THE\ CASE $\frac{\partial}{\partial y}\neq0,\frac{\partial}{\partial
x}=0$}

Using Equation (A56) and performing calculations of the corresponding
quantities, we find the components of the plasma dielectric permeability
tensor ($v_{i0y}$ has been changed by $u_{pl}$)%

\begin{align}
\varepsilon_{plxx}  & =\frac{\omega_{pi}^{2}}{\omega_{ci}^{2}}\left(
\frac{\partial}{\partial t}+u_{pl}\frac{\partial}{\partial y}\right)
^{2}\left(  \frac{\partial}{\partial t}\right)  ^{-2}\\
& -\frac{\omega_{pi}^{2}}{\omega_{ci}^{2}}\frac{1}{m_{i}}\left[  T_{i0}%
+T_{e0}-\frac{G_{1}+G_{3}}{D}\frac{\partial}{\partial t}-\frac{G_{2}+G_{4}}%
{D}\left(  \frac{\partial}{\partial t}+u_{pl}\frac{\partial}{\partial
y}\right)  \right]  \frac{\partial^{2}}{\partial y^{2}}\left(  \frac{\partial
}{\partial t}\right)  ^{-2},\nonumber\\
\varepsilon_{plxy}  & =\left(  \frac{\omega_{pi}^{2}\omega_{ci}}{\Omega
_{i}^{2}}+\frac{\omega_{pe}^{2}\omega_{ce}}{\Omega_{e}^{2}}\right)  \left(
\frac{\partial}{\partial t}\right)  ^{-1}\nonumber\\
& +\frac{\omega_{pi}^{2}}{\omega_{ci}^{3}}\frac{1}{m_{i}}\left[  T_{i0}%
-\frac{G_{2}+G_{4}}{D}\left(  \frac{\partial}{\partial t}+u_{pl}\frac
{\partial}{\partial y}\right)  \right]  \frac{\partial^{2}}{\partial y^{2}%
}\left(  \frac{\partial}{\partial t}\right)  ^{-1},\nonumber\\
\varepsilon_{plyx}  & =-\left(  \frac{\omega_{pi}^{2}\omega_{ci}}{\Omega
_{i}^{2}}+\frac{\omega_{pe}^{2}\omega_{ce}}{\Omega_{e}^{2}}\right)  \left(
\frac{\partial}{\partial t}\right)  ^{-1}\nonumber\\
& -\frac{\omega_{pi}^{2}}{\omega_{ci}^{3}}\frac{1}{m_{i}}\left[  T_{i0}%
-\frac{G_{3}}{D}\frac{\partial}{\partial t}-\frac{G_{4}}{D}\left(
\frac{\partial}{\partial t}+u_{pl}\frac{\partial}{\partial y}\right)  \right]
\frac{\partial^{2}}{\partial y^{2}}\left(  \frac{\partial}{\partial t}\right)
^{-1},\nonumber\\
\varepsilon_{plyy}  & =\frac{\omega_{pi}^{2}}{\omega_{ci}^{2}}.\nonumber
\end{align}
In obtaining expressions (21), we have taken into account that $m_{i}\gg
m_{e}$ and $n_{i0}\simeq n_{e0}$. Analogously from Equation (B21), we obtain
the cosmic ray dielectric permeability tensor%

\begin{align}
\varepsilon_{crxx}  & =\frac{\omega_{pcr}^{2}}{\omega_{ccr}^{2}}\gamma
_{cr0}^{3}\left(  \frac{\partial}{\partial t}+u_{cr}\frac{\partial}{\partial
y}\right)  ^{2}\left(  \frac{\partial}{\partial t}\right)  ^{-2}-\frac
{\omega_{pcr}^{2}}{\omega_{ccr}^{2}}\gamma_{cr0}^{2}c_{scr}^{2}\ \left(
\frac{u_{cr}}{c^{2}}\frac{\partial}{\partial t}+\frac{\partial}{\partial
y}\right)  \frac{\partial}{\partial y}\left(  \frac{\partial}{\partial
t}\right)  ^{-2},\\
\varepsilon_{crxy}  & =-\varepsilon_{cryx}=\frac{\omega_{pcr}^{2}}{\Omega
_{cr}^{2}}\omega_{ccr}\left(  \frac{\partial}{\partial t}\right)  ^{-1}%
+\frac{\omega_{pcr}^{2}}{\omega_{ccr}^{3}}\gamma_{cr0}^{3}c_{scr}^{2}\ \left(
\frac{u_{cr}}{c^{2}}\frac{\partial}{\partial t}+\frac{\partial}{\partial
y}\right)  \frac{\partial}{\partial y}\left(  \frac{\partial}{\partial
t}\right)  ^{-1},\nonumber\\
\varepsilon_{cryy}  & =\frac{\omega_{pcr}^{2}}{\omega_{ccr}^{2}}\gamma
_{cr0}.\nonumber
\end{align}
Here, we have used the additional condition for cosmic rays%
\[
1\gg\gamma_{cr0}^{3}\rho_{cr}^{2}\frac{u_{cr}}{c^{2}}\left(  \frac{\partial
}{\partial t}+u_{cr}\frac{\partial}{\partial y}\right)  \frac{\partial
}{\partial y}%
\]
(see Equation (B11)). The contribution of the term proportional to
$u_{cr}/c^{2}$ in Equation (22) is small. This term shows the contribution of
$\gamma_{cr1}$ to the cosmic ray pressure perturbation (see Equations (B8) and (B9)).

\bigskip

\subsection{Wave equation}

From Equations (17) and (18), using Equations (A54), (A55), (B19), and (B20)
and by omitting the contribution of the displacement current, we obtain the
equation
\begin{equation}
\varepsilon_{yy}c^{2}\left(  \frac{\partial}{\partial t}\right)  ^{-2}%
\frac{\partial^{2}E_{1x}}{\partial y^{2}}=\left(  \varepsilon_{xx}%
\varepsilon_{yy}-\varepsilon_{xy}\varepsilon_{yx}\right)  E_{1x},
\end{equation}
where $\varepsilon_{ij}=\varepsilon_{plij}+\varepsilon_{crij}$. The values
$\varepsilon_{ij}$ are defined by Equations (21) and (22). When calculating
the right-hand side of Equation (23), we assume some additional conditions
except those given by Equations (19) and (20). We will neglect the
contribution to $\varepsilon_{xy}\varepsilon_{yx}$ of the thermal cosmic ray
term in $\varepsilon_{crxy}$ and $\varepsilon_{cryx}$. Besides, we will use
the condition of quasineutrality in $\varepsilon_{xy}$ and $\varepsilon_{yx}$
and neglect the terms arising due to expansion of $\Omega_{i,cr}^{-2}$. An
analysis shows that the corresponding conditions can be written in the form
\begin{align}
& \max\left\{  \gamma_{cr0}\left(  \frac{\partial}{\partial t}+u_{cr}%
\frac{\partial}{\partial y}\right)  ^{2};c_{scr}^{2}\ \left(  \frac{u_{cr}%
}{c^{2}}\frac{\partial}{\partial t}+\frac{\partial}{\partial y}\right)
\frac{\partial}{\partial y}\right\} \\
& \gg\gamma_{cr0}^{3}\frac{c_{scr}^{4}}{\omega_{ccr}^{2}}\ \left(
\frac{u_{cr}}{c^{2}}\frac{\partial}{\partial t}+\frac{\partial}{\partial
y}\right)  ^{2}\frac{\partial^{2}}{\partial y^{2}};\gamma_{cr0}\frac
{c_{spl}^{2}c_{scr}^{2}}{\omega_{ci}\omega_{ccr}}\ \left(  \frac{u_{cr}}%
{c^{2}}\frac{\partial}{\partial t}+\frac{\partial}{\partial y}\right)
\frac{\partial^{3}}{\partial y^{3}};\nonumber\\
& \gamma_{cr0}\frac{c_{scr}^{2}}{\omega_{ci}\omega_{ccr}}\ \left(
\frac{u_{cr}}{c^{2}}\frac{\partial}{\partial t}+\frac{\partial}{\partial
y}\right)  \frac{\partial}{\partial y}\left(  \frac{\partial}{\partial
t}+u_{pl}\frac{\partial}{\partial y}\right)  ^{2},\nonumber
\end{align}
where $c_{spl}=\left(  2\gamma T_{i0}/m_{i}\right)  ^{1/2}$. According to
conditions (20) and (24), the contribution of the term $\varepsilon
_{xy}\varepsilon_{yx}$ to the Equation (23) is small. Thus, we obtain
\begin{equation}
c^{2}\frac{\partial^{2}E_{1x}}{\partial y^{2}}=\varepsilon_{xx}\left(
\frac{\partial}{\partial t}\right)  ^{2}E_{1x}.
\end{equation}

\bigskip

\subsection{Dispersion relation}

Using Equations (21) and (22) and accomplishing the Fourier transform in
Equation (25), we find for perturbations of the form $\exp\left(
ik_{y}y-i\omega t\right)  $ the following dispersion relation:
\begin{align}
0  & =\frac{\omega_{pi}^{2}}{\omega_{ci}^{2}}\left(  \omega-k_{y}%
u_{pl}\right)  ^{2}+\frac{\omega_{pcr}^{2}}{\omega_{ccr}^{2}}\gamma_{cr0}%
^{3}\left(  \omega-k_{y}u_{cr}\right)  ^{2}\\
& -\frac{\omega_{pi}^{2}}{\omega_{ci}^{2}}k_{y}^{2}\frac{1}{m_{i}}\left[
T_{i0}+T_{e0}+\frac{G_{1}+G_{3}}{D}i\omega+\frac{G_{2}+G_{4}}{D}i\left(
\omega-k_{y}u_{pl}\right)  \right] \nonumber\\
& -\frac{\omega_{pcr}^{2}}{\omega_{ccr}^{2}}\gamma_{cr0}^{2}k_{y}^{2}%
c_{scr}^{2}\ -k_{y}^{2}c^{2}.\nonumber
\end{align}
Below, we consider solutions of Equation (26) for the streaming instability
and an influence of the streaming and thermal pressure effects on the thermal instability.

\bigskip

\subsubsection{\textit{Streaming instability}}

Let us set all frequencies $\Omega$ equal to zero in Equation (26). To be more
specific, it means that $\omega-k_{y}u_{pl}\gg\Omega_{T,ni},\Omega_{\epsilon}$
and $\omega\gg\Omega_{T,ne},\Omega_{\epsilon}$, where $\Omega_{ie}\simeq
\Omega_{ei}=\Omega_{\epsilon}$ (the frequencies $\Omega$ are defined by
Equation (A12)). Then, this equation takes the form%
\begin{align}
0  & =\frac{\omega_{pi}^{2}}{\omega_{ci}^{2}}\left(  \omega-k_{y}%
u_{pl}\right)  ^{2}+\frac{\omega_{pcr}^{2}}{\omega_{ccr}^{2}}\gamma_{cr0}%
^{3}\left(  \omega-k_{y}u_{cr}\right)  ^{2}\\
& -\left(  \frac{\omega_{pi}^{2}}{\omega_{ci}^{2}}c_{spl}^{2}+\frac
{\omega_{pcr}^{2}}{\omega_{ccr}^{2}}\gamma_{cr0}^{2}c_{scr}^{2}+c^{2}\right)
k_{y}^{2}.\nonumber
\end{align}
The solution of Equation (27) is the following:
\begin{equation}
\omega=\frac{k_{y}\left(  u_{pl}+du_{cr}\right)  }{1+d}\pm\frac{k_{y}}%
{1+d}\left[  -\left(  u_{cr}-u_{pl}\right)  ^{2}d+\left(  1+d\right)  \left(
c_{spl}^{2}+\gamma_{cr0}^{-1}dc_{scr}^{2}+c_{Ai}^{2}\right)  \right]  ^{1/2},
\end{equation}
where
\begin{equation}
d=\frac{\omega_{ci}^{2}}{\omega_{pi}^{2}}\frac{\omega_{pcr}^{2}}{\omega
_{ccr}^{2}}\gamma_{cr0}^{3}=\frac{m_{cr}}{m_{i}}\frac{n_{cr0}}{n_{i0}}%
\gamma_{cr0}^{3}.
\end{equation}
We see that the streaming instability has a threshold $u_{crth}$ defined by
the sound and ion Alfv\'{e}n velocities%
\begin{equation}
u_{crth}^{2}=\left(  1+d^{-1}\right)  \left(  c_{spl}^{2}+\gamma_{cr0}%
^{-1}dc_{scr}^{2}+c_{Ai}^{2}\right)  .
\end{equation}
When this threshold is exceeded, $u_{cr}^{2}\gg u_{crth}^{2}$, the growth rate
$\delta_{gr}$ is given by%
\begin{equation}
\delta_{gr}=\frac{d^{1/2}}{1+d}k_{y}u_{cr}.
\end{equation}
These perturbations move with the phase velocity $v_{ph}=\left(
u_{pl}+du_{cr}\right)  /\left(  1+d\right)  $.

\bigskip

\subsubsection{\textit{Thermal instability}}

We now take into account the terms describing the thermal instability in
Equation (26). We consider the fast thermal energy exchange regime in which
$\Omega_{\epsilon}\gg\partial/\partial t,\Omega_{Ti,e},\Omega_{ni,e}$. Using
Equations (A29) and (A30), we will have%
\begin{equation}
\frac{\gamma\left(  2\omega-k_{y}u_{pl}\right)  +i\Omega_{T,n}}{\gamma\left(
2\omega-k_{y}u_{pl}\right)  +i\gamma\Omega_{T}}=c_{spl}^{-2}\left(
du_{cr}^{2}-\gamma_{cr0}^{-1}dc_{scr}^{2}-c_{Ai}^{2}+\frac{\omega^{2}}%
{k_{y}^{2}}\right)  ,
\end{equation}
where%
\begin{align*}
\Omega_{T,n}  & =\Omega_{Te}+\Omega_{Ti}-\Omega_{ne}-\Omega_{ni},\\
\Omega_{T}  & =\Omega_{Te}+\Omega_{Ti}.
\end{align*}
When obtaining Equation (32), we have assumed that $\omega\ll k_{y}u_{cr}$. If
the right-hand side of Equation (32) is much less than unity, we obtain
Field's isobaric solution $2\omega=k_{y}u_{pl}-i\Omega_{T,n}/\gamma$ (Field
1965). These perturbations travel with the phase velocity $u_{pl}/2$. In the
opposite case, Equation (32) has Parker's isochoric solution $2\omega
=k_{y}u_{pl}-i\Omega_{T}$ (Parker 1953). Thus, the presence of streaming
cosmic rays can change the kind of thermal instability. When the right-hand
side of Equation (32) is of the order of unity, the limiting solutions intermix.

\bigskip

\section{THE\ CASE\ $\frac{\partial}{\partial x}\neq0,\frac{\partial}{\partial
y}=0$}

Calculating the components of the plasma dielectric permeability tensor given
by Equation (A56), we obtain%
\begin{align}
\varepsilon_{plxx}  & =\frac{\omega_{pi}^{2}}{\omega_{ci}^{2}},\\
\varepsilon_{plxy}  & =\frac{\omega_{pi}^{2}\omega_{ci}}{\Omega_{i}^{2}%
}\left(  \frac{\partial}{\partial t}\right)  ^{-1}+\frac{\omega_{pe}^{2}%
\omega_{ce}}{\Omega_{e}^{2}}\left(  \frac{\partial}{\partial t}\right)
^{-1}\nonumber\\
& +\frac{\omega_{pi}^{2}}{\omega_{ci}^{3}}\left[  \frac{1}{m_{i}}\left(
T_{i0}-\frac{G_{3}+G_{4}}{D}\frac{\partial}{\partial t}\right)  \frac
{\partial}{\partial x}-\omega_{ci}u_{pl}\right]  \frac{\partial}{\partial
x}\left(  \frac{\partial}{\partial t}\right)  ^{-1},\nonumber\\
\varepsilon_{plyx}  & =-\frac{\omega_{pi}^{2}\omega_{ci}}{\Omega_{i}^{2}%
}\left(  \frac{\partial}{\partial t}\right)  ^{-1}-\frac{\omega_{pe}^{2}%
\omega_{ce}}{\Omega_{e}^{2}}\left(  \frac{\partial}{\partial t}\right)
^{-1}\nonumber\\
& -\frac{\omega_{pi}^{2}}{\omega_{ci}^{3}}\left[  \frac{1}{m_{i}}\left(
T_{i0}-\frac{G_{2}+G_{4}}{D}\frac{\partial}{\partial t}\right)  \frac
{\partial}{\partial x}+\omega_{ci}u_{pl}\right]  \frac{\partial}{\partial
x}\left(  \frac{\partial}{\partial t}\right)  ^{-1},\nonumber\\
\varepsilon_{plyy}  & =\frac{\omega_{pi}^{2}}{\omega_{ci}^{2}}-\frac
{\omega_{pi}^{2}}{\omega_{ci}^{2}}\frac{1}{m_{i}}\left(  T_{i0}+T_{e0}%
-\frac{G_{1}+G_{2}+G_{3}+G_{4}}{D}\frac{\partial}{\partial t}\right)
\frac{\partial^{2}}{\partial x^{2}}\left(  \frac{\partial}{\partial t}\right)
^{-2}\nonumber\\
& +\frac{\omega_{pi}^{2}}{\omega_{ci}^{2}}u_{pl}^{2}\frac{\partial^{2}%
}{\partial x^{2}}\left(  \frac{\partial}{\partial t}\right)  ^{-2}%
-\frac{\omega_{pi}^{2}}{\omega_{ci}^{3}}u_{pl}\frac{1}{m_{i}}\frac{G_{2}%
-G_{3}}{D}\frac{\partial^{3}}{\partial x^{3}}\left(  \frac{\partial}{\partial
t}\right)  ^{-1}.\nonumber
\end{align}
From Equation (B21) for cosmic rays, we will have
\begin{align}
\varepsilon_{crxx}  & =\frac{\omega_{pcr}^{2}}{\omega_{ccr}^{2}}\gamma
_{cr0}^{3},\\
\varepsilon_{crxy}  & =\frac{\omega_{pcr}^{2}\omega_{ccr}}{\Omega_{cr}^{2}%
}\left(  \frac{\partial}{\partial t}\right)  ^{-1}+\frac{\omega_{pcr}^{2}%
}{\omega_{ccr}^{3}}\gamma_{cr0}^{3}\left(  \gamma_{cr0}^{2}c_{scr}^{2}%
\frac{\partial}{\partial x}-\omega_{ccr}u_{cr}\right)  \frac{\partial
}{\partial x}\left(  \frac{\partial}{\partial t}\right)  ^{-1},\nonumber\\
\varepsilon_{cryx}  & =-\frac{\omega_{pcr}^{2}\omega_{ccr}}{\Omega_{cr}^{2}%
}\left(  \frac{\partial}{\partial t}\right)  ^{-1}-\frac{\omega_{pcr}^{2}%
}{\omega_{ccr}^{3}}\gamma_{cr0}^{3}\left(  c_{scr}^{2}\frac{\partial}{\partial
x}+\omega_{ccr}u_{cr}\right)  \frac{\partial}{\partial x}\left(
\frac{\partial}{\partial t}\right)  ^{-1},\nonumber\\
\varepsilon_{cryy}  & =\frac{\omega_{pcr}^{2}}{\omega_{ccr}^{2}}\gamma
_{cr0}\left[  1+\gamma_{cr0}^{2}u_{cr}^{2}\frac{\partial^{2}}{\partial x^{2}%
}\left(  \frac{\partial}{\partial t}\right)  ^{-2}\right]  -\frac{\omega
_{pcr}^{2}}{\omega_{ccr}^{2}}c_{scr}^{2}\gamma_{cr0}^{2}\frac{\partial^{2}%
}{\partial x^{2}}\left(  \frac{\partial}{\partial t}\right)  ^{-2}.\nonumber
\end{align}
In this case, the additional condition for cosmic rays except for Equation
(20) is the following:
\[
1\gg\gamma_{cr0}^{2}\frac{u_{cr}c_{scr}}{c^{2}}\rho_{cr}\frac{\partial
}{\partial x}%
\]
(see Equation (B11)).

\bigskip

\subsection{Wave equation}

In the case under consideration, the wave equation has the form%
\begin{equation}
\varepsilon_{xx}c^{2}\left(  \frac{\partial}{\partial t}\right)  ^{-2}%
\frac{\partial^{2}E_{1y}}{\partial x^{2}}=\left(  \varepsilon_{xx}%
\varepsilon_{yy}-\varepsilon_{xy}\varepsilon_{yx}\right)  E_{1y}.
\end{equation}
Using Equations (33) and (34) and calculating the right-hand side of Equation
(35), we find
\begin{align}
\left(  \varepsilon_{xx}\varepsilon_{yy}-\varepsilon_{xy}\varepsilon
_{yx}\right)   & =\varepsilon_{xx}\left(  \frac{\omega_{pi}^{2}}{\omega
_{ci}^{2}}+\frac{\omega_{pcr}^{2}}{\omega_{ccr}^{2}}\gamma_{cr0}\right)
-\varepsilon_{xx}\frac{\omega_{pcr}^{2}}{\omega_{ccr}^{2}}\gamma_{cr0}%
^{2}c_{scr}^{2}\frac{\partial^{2}}{\partial x^{2}}\left(  \frac{\partial
}{\partial t}\right)  ^{-2}\\
& -\varepsilon_{xx}\frac{\omega_{pi}^{2}}{\omega_{ci}^{2}}\frac{1}{m_{i}%
}\left(  T_{i0}+T_{e0}-\frac{G_{1}+G_{2}+G_{3}+G_{4}}{D}\frac{\partial
}{\partial t}\right)  \frac{\partial^{2}}{\partial x^{2}}\left(
\frac{\partial}{\partial t}\right)  ^{-2}\nonumber\\
& +\frac{\omega_{pi}^{2}}{\omega_{ci}^{2}}\frac{\omega_{pcr}^{2}}{\omega
_{ccr}^{2}}\gamma_{cr0}^{3}\left(  u_{cr}-u_{pl}\right)  ^{2}\frac
{\partial^{2}}{\partial x^{2}}\left(  \frac{\partial}{\partial t}\right)
^{-2}.\nonumber
\end{align}

\bigskip

\subsection{Dispersion relation}

After Fourier transformation of Equation (35) and substitution of Equation
(36), we derive the dispersion relation%
\begin{align}
\left(  1+\frac{\omega_{ci}^{2}}{\omega_{pi}^{2}}\frac{\omega_{pcr}^{2}%
}{\omega_{ccr}^{2}}\gamma_{cr0}\right)  \omega^{2}  & =k_{x}^{2}c_{Ai}%
^{2}+\frac{\omega_{ci}^{2}}{\omega_{pi}^{2}}\frac{\omega_{pcr}^{2}}%
{\omega_{ccr}^{2}}\gamma_{cr0}^{2}k_{x}^{2}c_{scr}^{2}\\
& +k_{x}^{2}\frac{1}{m_{i}}\left(  T_{i0}+T_{e0}+\frac{G_{1}+G_{2}+G_{3}%
+G_{4}}{D}i\omega\right) \nonumber\\
& -\frac{1}{\varepsilon_{xx}}\frac{\omega_{pcr}^{2}}{\omega_{ccr}^{2}}%
\gamma_{cr0}^{3}k_{x}^{2}\left(  u_{cr}-u_{pl}\right)  ^{2}.\nonumber
\end{align}
Below, as above, we consider the streaming instability and influence of cosmic
rays on the thermal instability.

\bigskip

\subsubsection{\textit{Streaming instability}}

As above, we again neglect in the values $G_{i},i=1,2,3,4,$ and $D$ all the
frequencies $\Omega$. Then, Equation (37) takes the form%
\begin{equation}
\left(  1+\gamma_{cr0}^{-2}d\right)  \frac{\omega^{2}}{k_{x}^{2}}=-\frac
{d}{\left(  1+d\right)  }u_{cr}^{2}+c_{spl}^{2}+\gamma_{cr0}^{-1}dc_{scr}%
^{2}+c_{Ai}^{2},
\end{equation}
where we have omitted $u_{pl}$ in comparison with $u_{cr}$. This equation
describes an aperiodic instability, if the drift velocity of cosmic rays
exceeds the threshold value given by Equation (30). An expression for the
growth rate $\delta_{gr}$ when $u_{cr}$ exceeds $u_{crth}$ is the following:%
\begin{equation}
\delta_{gr}=\left[  \frac{d}{\left(  1+d\right)  \left(  1+\gamma_{cr0}%
^{-2}d\right)  }\right]  ^{1/2}k_{x}u_{cr}.
\end{equation}

\bigskip

\subsubsection{\textit{Thermal instability}}

Now, we take into account the contribution into Equation (37) of terms
describing the thermal instability in the fast thermal energy exchange regime
$\Omega_{\epsilon}\gg\partial/\partial t,\Omega_{Ti,e},\Omega_{ni,e}$. The
dispersion relation becomes%
\begin{equation}
\frac{2\gamma\omega+i\Omega_{T,n}}{2\gamma\omega+i\gamma\Omega_{T}}%
=c_{spl}^{-2}\left[  \frac{d}{\left(  1+d\right)  }u_{cr}^{2}-c_{Ai}%
^{2}-\gamma_{cr0}^{-1}dc_{scr}^{2}+\left(  1+\gamma_{cr0}^{-2}d\right)
\frac{\omega^{2}}{k_{x}^{2}}\right]  .
\end{equation}
This equation is analogous to Equation (32). Depending on whether the
right-hand side of Equation (40) is much larger than the unity or not, we will
have Parker's or Field's instability (see above). In these limiting cases, the
value $\omega^{2}$ on the right-hand side of Equation (40) must be substituted
by $-\Omega_{T}^{2}$ or $-\Omega_{T,n}^{2}$, respectively.

\bigskip

\section{DISCUSSION AND\ IMPLICATIONS}

The growth rates (31) and (39) of streaming instabilities have a similar form
and increase with decreasing of the perturbation wavelength. The thresholds
for the cases $k_{x}=0,k_{y}\neq0$ and $k_{x}\neq0,k_{y}=0$ are equal to each
other (see Equations (28) at $u_{cr}\gg u_{pl}$ and (38)). Thus, streaming
cosmic rays generate perturbations in all directions across the ambient
magnetic field. A spectrum of the perturbations in the $\mathbf{k}$-space is
limited from above by conditions given by Equations (19) and (20) and
additional conditions (see inequalities after Equations (22) and (34)). These
conditions for the case $k_{x}=0,k_{y}\neq0$ can be written in the form%
\[
\left(  \frac{\lambda_{y}}{2\pi}\right)  ^{2}\gg\max\left\{  \frac{d}%
{1+d}\frac{u_{cr}^{2}}{\omega_{ci}^{2}};\frac{\gamma_{cr0}^{4}}{\left(
1+d\right)  }\frac{u_{cr}^{2}}{\omega_{ccr}^{2}};\frac{\gamma_{cr0}^{3}%
}{\left(  1+d\right)  ^{1/2}}\frac{u_{cr}^{2}}{\omega_{ccr}^{2}}\frac
{c_{scr}^{2}}{c^{2}}\right\}  ,
\]
where the value $d$ is defined by Equation (29) and $\lambda$ is the
wavelength. We have assumed that the threshold of instability is exceeded. The
conditions (20) are satisfied. The analogous conditions for the case
$k_{x}\neq0,k_{y}=0$ are the following:%
\[
\left(  \frac{\lambda_{x}}{2\pi}\right)  ^{2}\gg\max\left\{  \frac{d}{\left(
1+d\right)  \left(  1+\gamma_{cr0}^{-2}d\right)  }\frac{u_{cr}^{2}}%
{\omega_{ci}^{2}};\frac{\gamma_{cr0}^{4}d}{\left(  1+d\right)  \left(
1+\gamma_{cr0}^{-2}d\right)  }\frac{u_{cr}^{2}}{\omega_{ccr}^{2}};\rho_{i}%
^{2};\gamma_{cr0}^{3}\rho_{cr}^{2};\gamma_{cr0}^{4}\frac{u_{cr}^{2}}%
{\omega_{ccr}^{2}}\frac{c_{scr}^{4}}{c^{4}}\right\}  .
\]

Let us consider the polarization of perturbations. In the case $k_{x}=0,$
$k_{y}\neq0$, the current $j_{1y}=0$ (see Equation (18) without the
displacement current). Then, the component of the electric field $E_{1y}$ is
equal to $E_{1y}=-\left(  \varepsilon_{yx}/\varepsilon_{yy}\right)  E_{1x}$.
Estimations show that $\left(  \varepsilon_{yx}/\varepsilon_{yy}\right)  \ll1$
for the streaming instability. Thus, the polarization is a linear one being
$E_{1x}\gg E_{1y}$. The electric field polarization for the thermal
instability depends on the wavelength of perturbations, ion Alfv\'{e}n
velocity, and parameters of cosmic rays and can be various. In the case
$k_{x}\neq0,$ $k_{y}=0$, the current $j_{1x}=0$ (see Equation (17)). Then
$E_{1x}=-\left(  \varepsilon_{xy}/\varepsilon_{xx}\right)  E_{1y}$. The ratio
$\varepsilon_{xy}/\varepsilon_{xx}$ for the streaming instability is given by%
\[
\frac{\varepsilon_{xy}}{\varepsilon_{xx}}=-if\left[  \frac{d\left(
1+\gamma_{cr0}^{-2}d\right)  }{\left(  1+d\right)  }\right]  ^{1/2}%
+\text{terms}\ll1,
\]
where $f=1-q_{cr}m_{i}/\gamma_{cr0}^{3}q_{i}m_{cr}$. We have assumed that
\[
\frac{d}{\left(  1+d\right)  }\gamma_{cr0}^{2}\frac{k_{x}^{2}c_{scr}^{2}%
}{\omega_{ccr}\delta_{gr}}\ll1,
\]
where $\delta_{gr}$ is determined by Equation (39). This condition can be
easily satisfied taking into account Equation (20). Thus, we see that the
ratio $\varepsilon_{xy}/\varepsilon_{xx}$ can be both smaller and larger than
the unity. This result is also just for the case of thermal instability.

From Equations (32) and (40), it is followed that the relations between
hydrodynamical parameters of thermal plasma and cosmic rays and the
perturbation wavelength determine the kind of thermal instability from
Parker's(1953) to Field's (1965) type instability. Dissipative processes such
as the thermal conductivity of plasma and cosmic rays can affect the growth
rate of thermal instability.

We now compare the growth rate found for the streaming instability along the
background magnetic field (Nekrasov \& Shadmehri 2012) with the growth rates
obtained in this paper. The growth rates given by Equations (31) and (39) are
of the same order of magnitude, if $\gamma_{cr0}\sim1$ or $\gamma_{cr0}\gg1$
and $d\lesssim1$ (for the same wavenumbers). In the case $\gamma_{cr0}\gg1$
and $d\gg1$, the growth rate given by Equation (39) is larger. Therefore, we
use Equation (39) for a comparison. The maximal growth rate found by Nekrasov
\& Shadmehri (2012) is equal to%
\[
\delta_{m}=2j_{cr0}\left(  \frac{\pi}{m_{cr}n_{cr0}c^{2}}\right)
^{1/2}\left(  \frac{\gamma_{cr0}^{-1}c_{A}^{2}}{\gamma_{cr0}^{-1}c_{scr}%
^{2}+c_{A}^{2}}\right)  ^{1/2},
\]
where $c_{A}=c_{Ai}\left(  1+\gamma_{cr0}^{-2}d\right)  ^{-1/2}$. The ratio of
this growth rate to the growth rate (39) for the same cosmic ray drift
velocities is the following:%
\[
\frac{\delta_{m}}{\delta_{gr}}=\left(  1+d^{-1}\right)  ^{1/2}\frac{c_{Ai}%
}{\left(  c_{scr}^{2}+\gamma_{cr0}c_{A}^{2}\right)  ^{1/2}}\frac{\omega_{pcr}%
}{k_{x}c}.
\]
We see that for sufficiently short wavelengths the ratio $\delta_{m}%
/\delta_{gr}$ can be less than unity. Thus, the transverse streaming
instabilities induced by cosmic rays can considerably contribute to turbulence
of astrophysical objects and amplification of magnetic fields.

We have explored the situation in which cosmic rays drift across the
background magnetic field. This model has been considered by Riquelme \&
Spitkovsky (2010) for the problem of the magnetic field amplification in the
upstream region of the supernova remnant shocks. The perturbations along the
background magnetic field have been investigated and the cosmic ray
back-reaction has not been taken into account in the analytical treatment. The
latter effect for the longitudinal perturbations has been included in the
paper by Nekrasov \& Shadmehri (2012) where the growth rate considerably
larger than that in (Riquelme \& Spitkovsky 2010) has been found. In this
paper, we have investigated the transverse perturbations. In the paper by Bell
(2005), the unstable perturbations for the last case are absent in the MHD
model. However, the multi-fluid approach gives a different result. In another
model, cosmic rays drift along the magnetic field. This case has been
investigated by Bell (2004) (see also Riquelme \& Spitkovsky 2009). In both
cases (Bell 2004; Riquelme \& Spitkovsky 2010), the growth rates have turned
out to be the same.

The streaming cosmic ray driven instabilities can exist in a variety of
environments. Although such a type of instability was suggested originally for
the magnetic field amplification in the shocks of supernovae, we think,
wherever there is a strong cosmic ray streaming, this instability may play a
significant role. For example, the models described above can be applied to
the ICM where cosmic rays are an important ingredient (Loewenstein et al.
1991; Guo \& Oh 2008; Sharma et al. 2009; Sharma, Parrish \& Quataert 2010).
Observations show that many cavities or bubbles in the ICM contain cosmic rays
and magnetic field (e.g. Guo \& Oh 2008). A substantial amount of cosmic rays
may escape from these buoyantly rising bubbles (e.g. En\ss lin 2003) which
could be shredded or disrupted by RT and KH instabilities as they rise through
the ICM (e.g. Fabian et al. 2006). Cosmic rays may also be produced by other
processes near the central AGN of the galaxy cluster. Structure formation
shocks, merger shocks and supernovae may also inject cosmic rays into the ICM
(e.g. Voelk, Aharonian \& Breitschwerdt 1996; Berezinsky, Blasi \& Ptuskin
1997). The observation of diffuse radio synchrotron emission in many galaxy
clusters give direct evidence for the presence of an extensive population of
non-thermal particles (e.g. Brunetti et al. 2001; Pfrommer \& En\ss lin 2004).
Recent Chandra and XMM observations also show evidence for a significant
non-thermal particle population within the ICM (Sanders, Fabian \& Dunn 2005;
Werner et al. 2007).

In some of supernova remnants such as IC 443, SN 1006, Kepler, Tycho and etc.,
the driven shocks are propagating in a partially ionized ambient medium. This
was a good motivation to extend cosmic ray streaming instability from the MHD
approach to a two-fluid case, by considering ions and neutrals as two separate
fluids where they can exchange momentum via collisions (e.g. Reville et al.
2007; see also Bykov \& Toptygin 2005). It has been shown that the instability
is getting slower rate because of collisions of ions with neutrals, in
particular when the cosmic ray flux is not very strong. However, the
back-reaction of cosmic rays has not been considered. Having in mind the
finding that the growth rate is significantly enhanced in the presence of
cosmic ray back-reaction in a three-fluid plasma system consisting of the
ions, electrons, and cosmic rays,\ one may naturally expect such an effect in
a four-fluid plasma system consisting of the ions, electrons, cosmic rays, and
neutrals. It deserves a further study, but we may expect that the stabilizing
effect of the ion-neutral collisions can be compensated by the back-reaction
of cosmic rays.

\bigskip

\section{CONCLUSION}

Using the multi-fluid approach, we have investigated streaming and thermal
instabilities of the electron-ion plasma with homogeneous cold cosmic rays
drifting across the background magnetic field. We have taken into account the
return current of the background plasma and the back-reaction of cosmic rays
for perturbations transverse to the magnetic field and along and across to the
cosmic ray drift velocity. For sufficiently short wavelength perturbations,
the growth rates exceed the one of streaming instability along the magnetic field.

The thermal instability has been shown not to be subjected to the action of
cosmic rays in the model under consideration. The dispersion relations for the
thermal instability in the multi-fluid approach has been derived which include
sound velocities of plasma and cosmic rays, Alfv\'{e}n and cosmic ray drift
velocities. The relations between these parameters determine the kind of
thermal instability from Parker's to Field's type instability.

The results of this paper can be useful for the investigation of the
electron-ion astrophysical objects such as galaxy clusters including the
dynamics of streaming cosmic rays.

\bigskip

\section{REFERENCES}

Achterberg A., 1983, A\&A, 119, 274

Alfv\'{e}n H., 1939, Phys. Rev., 55, 425

Balbus S. A., Soker N., 1989, ApJ, 341, 611

Begelman M. C., \& McKee C. F., 1990, ApJ, 358, 375

Begelman M. C., Zweibel E. G., 1994, ApJ, 431, 689

Bell A. R., 2004, MNRAS, 353, 550

Bell A. R., 2005, MNRAS, 358, 181

Berezinsky V. S., Blasi P., Ptuskin V. S., 1997, ApJ, 487, 529

Beresnyak A., Jones T. W., Lazarian A., 2009, ApJ, 707, 1541

Bogdanovi\'{c} T., Reynolds C. S., Balbus S. A., Parrish I. J., 2009, ApJ,
704, 211

Braginskii S. I., 1965, Rev. Plasma Phys., 1, 205

Brunetti G., Setti G., Feretti L., Giovannini G., 2001, MNRAS, 320, 365

Bykov A. M., Toptygin I. N., 2005, Astronomy Letters, 31, 839

Cavagnolo K. W., Donahue M., Voil G. M., Sun M., 2008, ApJ, 683, L107

Conselice C. J., Gallagher J. S., III, Wyse R. F. G., 2001, AJ, 122, 2281

Dzhavakhishvili D. I., Tsintsadze N. L., 1973, Sov. Phys. JEPT, 37, 666; 1973,
Zh. Eksp. Teor. Fiz. 64, 1314

En\ss lin T., Pfrommer C., Miniati F., Subramanian K., 2011, A\&A, 527, A99

Everett J. E., Zweibel E. G., Benjamin R. A., McCammon D., Rocks L., Gallagher
J. S., III, 2008, ApJ, 674, 258

Fabian A. C., Sanders J. S., Taylor G. B., Allen S. W., Crawford C. S.,
Johnstone R. M., Iwasawa K., 2006, MNRAS, 366, 417

Ferland G. J., Fabian A. C., Hatch N. A., Johnstone R. M., Porter R. L., van
Hoof P. A. M., Williams R. J. R., 2009, MNRAS, 392, 1475

Field G.B., 1965, ApJ, 142, 531

Field G. B., Goldsmith D. W., Habing H. J., 1969, ApJ, 155, L149

Fukue T., Kamaya H., 2007, ApJ, 669, 363

Gammie C. F., 1996, ApJ, 457, 355

Guo F., Oh S. P., 2008, MNRAS, 384, 251

Hennebelle P., P\'{e}rault M., 2000, A\&A, 359, 1124

Inoue T., Inutsuka S., 2008, ApJ, 687, 303

Koyama H., Inutsuka S., 2000, ApJ, 532, 980

Kulsrud R., Pearce W. P., 1969, ApJ, 156, 445

Loewenstein M., 1990, ApJ, 349, 471

Loewenstein M., Zweibel E. G., Begelman M. C., 1991, ApJ, 377, 392

Lontano M., Bulanov S., Koga J., 2002, AIP Conf. Proc., 611, 157

Mathews W., Bregman J., 1978, ApJ, 224, 308

Nekrasov A. K., 2007, Phys. Plasmas, 14, 062107

Nekrasov A. K., 2011, ApJ, 739, 88

Nekrasov A. K., 2012, MNRAS, 419, 522

Nekrasov A. K., Shadmehri M., 2010, ApJ, 724,\textbf{\ }1165

Nekrasov A. K., Shadmehri M., 2011, Astrophys. Space Sci., 333, 477

Nekrasov A. K., Shadmehri M., 2012, Astro-ph., arXiv:1203.5734 (accepted by ApJ)

O'Dea C. P. et al., 2008, ApJ, 681, 1035

Parker E. N., 1953, ApJ, 117, 431

Parrish I. J., Quataert E., Sharma P., 2009, ApJ, 703, 96

Pfrommer C., En\ss lin T. A., 2004, A\&A, 413, 17

Reville B., Kirk J. G., Duffy P., O'Sullivan S., 2007, A\&A, 475, 435

Riquelme M. A., Spitkovsky A., 2009, ApJ, 694, 626

Riquelme M. A., Spitkovsky A. 2010, ApJ, 717, 1054

Salom\'{e} P. et al., 2006, A\&A, 454, 437

Samui S., Subramanian K., Srianand R., 2010, MNRAS, 402, 2778

S\'{a}nchez-Salcedo F. J., V\'{a}zquez-Semadeni E., Gazol A., 2002, ApJ, 577, 768

Sanders J. S., Fabian A. C., Dunn R. J. H., 2005, MNRAS, 360, 133

Shadmehri M., Nejad-Asghar M., Khesali A., 2010, Ap\&SS, 326, 83

Sharma P., Chandran B. D. G., Quataert E., Parrish I. J., 2009, ApJ, 699, 348

Sharma P., Parrish I. J., Quataert E., 2010, ApJ, 720, 652

Toepfer A. J., 1971, Phys. Rev. A, 3, 1444

Tozzi P., Norman C., 2001, ApJ, 546, 63

V\'{a}zquez-Semadeni E., Ryu D., Passot T., Gonz\'{a}lez R. F., Gazol A.,
2006, ApJ, 643, 245

Voelk H. J., Aharonian F. A., Breitschwerdt D., 1996, Space Sci. Rev., 75, 279

Werner N., Kaastra J. S., Takei Y., Lieu R., Vink J., Tamura T., 2007, A\&A,
468, 849

Yusef-Zadeh F., Wardle M., Roy S., 2007, ApJ, 665, L123

Zweibel E. G., 2003, ApJ, 587, 625

Zweibel E. G., Everett J. E., 2010, ApJ, 709, 1412

\bigskip
\begin{appendix}

\section{Appendix}

\subsection{Perturbed velocities of ions and electrons}

We put in Equation (1) $\mathbf{v}_{j}=\mathbf{v}_{j0}+\mathbf{v}_{j1}$,
$p_{j}=p_{j0}+p_{j1}$, $\mathbf{E=E}_{0}+\mathbf{E}_{1}$, $\mathbf{B=B}%
_{0}+\mathbf{B}_{1}$, where the subscript $0$ denotes equilibrium uniform
parameters and the subscript $1$ relates to perturbations. Then the linearized
version of this equation takes the form%
\begin{equation}
\frac{\partial\mathbf{v}_{j1}}{\partial t}+\mathbf{v}_{j0}\cdot\mathbf{\nabla
v}_{j1}=-\frac{\mathbf{\nabla}T_{j1}}{m_{j}}-\frac{T_{j0}}{m_{j}}%
\frac{\mathbf{\nabla}n_{j1}}{n_{j0}}+\mathbf{F}_{j1}\mathbf{+}\frac{q_{j}%
}{m_{j}c}\mathbf{v}_{j1}\times\mathbf{B}_{0},\tag{A1}%
\end{equation}
where we have used that $p_{j1}=n_{j0}T_{j1}+n_{j1}T_{j0}$ ($n_{j}%
=n_{j0}+n_{j1}$, $T_{j}=T_{j0}+T_{j1}$) and introduced notation%
\begin{equation}
\mathbf{F}_{j1}=\frac{q_{j}}{m_{j}}\mathbf{E}_{1}\mathbf{+}\frac{q_{j}}%
{m_{j}c}\mathbf{v}_{j0}\times\mathbf{B}_{1}.\tag{A2}%
\end{equation}
From Equation (A1), we find expressions for the ion velocities $v_{i1x,y}$ in
the form%
\begin{align}
\Omega_{i}^{2}v_{i1x}  & =\frac{1}{m_{i}}L_{ix}T_{i1}-\frac{T_{i0}}{m_{i}%
}L_{ix}\left(  \frac{\partial}{\partial t}+v_{i0y}\frac{\partial}{\partial
y}\right)  ^{-1}\mathbf{\nabla}\cdot\mathbf{v}_{i1}\tag{A3}\\
& \mathbf{+}\omega_{ci}F_{i1y}+\left(  \frac{\partial}{\partial t}%
+v_{i0y}\frac{\partial}{\partial y}\right)  F_{i1x}\nonumber
\end{align}
and
\begin{align}
\Omega_{i}^{2}v_{i1y}  & =\frac{1}{m_{i}}L_{iy}T_{i1}-\frac{T_{i0}}{m_{i}%
}L_{iy}\left(  \frac{\partial}{\partial t}+v_{i0y}\frac{\partial}{\partial
y}\right)  ^{-1}\mathbf{\nabla}\cdot\mathbf{v}_{i1}\tag{A4}\\
& -\omega_{ci}F_{i1x}+\left(  \frac{\partial}{\partial t}+v_{i0y}%
\frac{\partial}{\partial y}\right)  F_{i1y}.\nonumber
\end{align}
In Equations (A3) and (A4), we have used the linearized continuity equation
(2). The following notations are here introduced:%
\begin{align}
\Omega_{i}^{2}  & =\left(  \frac{\partial}{\partial t}+v_{i0y}\frac{\partial
}{\partial y}\right)  ^{2}+\omega_{ci}^{2},\tag{A5}\\
L_{ix}  & =-\omega_{ci}\frac{\partial}{\partial y}-\left(  \frac{\partial
}{\partial t}+v_{i0y}\frac{\partial}{\partial y}\right)  \frac{\partial
}{\partial x},\nonumber\\
L_{iy}  & =\omega_{ci}\frac{\partial}{\partial x}-\left(  \frac{\partial
}{\partial t}+v_{i0y}\frac{\partial}{\partial y}\right)  \frac{\partial
}{\partial y}.\nonumber
\end{align}
Analogous equations for the electrons are the following:%
\begin{equation}
\Omega_{e}^{2}v_{e1x}=\frac{1}{m_{e}}L_{ex}T_{e1}-\frac{T_{e0}}{m_{e}}%
L_{ex}\left(  \frac{\partial}{\partial t}\right)  ^{-1}\mathbf{\nabla}%
\cdot\mathbf{v}_{e1}+\omega_{ce}F_{e1y}+\frac{\partial F_{e1x}}{\partial
t},\tag{A6}%
\end{equation}%
\begin{equation}
\Omega_{e}^{2}v_{e1y}=\frac{1}{m_{e}}L_{ey}T_{e1}-\frac{T_{e0}}{m_{e}}%
L_{ey}\left(  \frac{\partial}{\partial t}\right)  ^{-1}\mathbf{\nabla}%
\cdot\mathbf{v}_{e1}-\omega_{ce}F_{e1x}+\frac{\partial F_{e1y}}{\partial
t},\tag{A7}%
\end{equation}
where
\begin{align}
\Omega_{e}^{2}  & =\frac{\partial^{2}}{\partial t^{2}}+\omega_{ce}%
^{2},\tag{A8}\\
L_{ex}  & =-\omega_{ce}\frac{\partial}{\partial y}-\frac{\partial^{2}%
}{\partial x\partial t},\nonumber\\
L_{ey}  & =\omega_{ce}\frac{\partial}{\partial x}-\frac{\partial^{2}}{\partial
y\partial t}.\nonumber
\end{align}
We do not consider the longitudinal velocity $v_{j1z}$ because as can be shown
in the case $\partial/\partial z=0$ this velocity only depends on the electric
field $E_{1z}$, $\partial v_{j1z}/\partial t=\left(  q_{j}/m_{j}\right)
E_{1z}$, and the transverse and longitudinal wave equations are split.

\bigskip

\subsection{Perturbed temperatures of ions and electrons\textit{\ }}

We find now equations for the temperature perturbations $T_{i,e1}$. We here
assume that equilibrium temperatures $T_{i0}$ and $T_{e0}$ are equal one
another, $T_{i0}=T_{e0}=T_{0}$. The case $T_{i0}\neq T_{e0}$ for thermal
instability has been considered by Nekrasov (2011, 2012). For equal
temperatures, the terms connected with the perturbation of thermal energy
exchange frequency in Equations (3) and (4) will be absent. However for
convenience of calculations, we formally retain different notations for the
ion and electron temperatures. From Equations (3) and (4) in the linear form,
we obtain equations for the temperature perturbations
\begin{equation}
D_{1i}T_{i1}-D_{2i}T_{e1}=C_{1i}\mathbf{\nabla}\cdot\mathbf{v}_{i1},\tag{A9}%
\end{equation}%
\begin{equation}
D_{1e}T_{e1}-D_{2e}T_{i1}=C_{1e}\mathbf{\nabla}\cdot\mathbf{v}_{e1},\tag{A10}%
\end{equation}
where notations are introduced%
\begin{align}
D_{1i}  & =\left[  \left(  \frac{\partial}{\partial t}+v_{i0y}\frac{\partial
}{\partial y}\right)  +\Omega_{Ti}+\Omega_{ie}\right]  \left(  \frac{\partial
}{\partial t}+v_{i0y}\frac{\partial}{\partial y}\right)  ,\tag{A11}\\
D_{2i}  & =\Omega_{ie}\left(  \frac{\partial}{\partial t}+v_{i0y}%
\frac{\partial}{\partial y}\right)  ,\nonumber\\
C_{1i}  & =T_{i0}\left[  -\left(  \gamma-1\right)  \left(  \frac{\partial
}{\partial t}+v_{i0y}\frac{\partial}{\partial y}\right)  +\Omega_{ni}\right]
,\nonumber\\
D_{1e}  & =\left(  \frac{\partial}{\partial t}+\Omega_{Te}+\Omega_{ei}\right)
\frac{\partial}{\partial t},\nonumber\\
D_{2e}  & =\Omega_{ei}\frac{\partial}{\partial t},\nonumber\\
C_{1e}  & =T_{e0}\left[  -\left(  \gamma-1\right)  \frac{\partial}{\partial
t}+\Omega_{ne}\right]  .\nonumber
\end{align}
When obtaining Equations (A9) and (A10), we have used Equations (2) and (16).
The frequencies in Equation (A11) are the following:%
\begin{align}
\Omega_{Tj}  & =\left(  \gamma-1\right)  \frac{\partial%
\mathcal{L}%
_{j}\left(  n_{j0},T_{j0}\right)  }{n_{j0}\partial T_{j0}},\Omega_{nj}=\left(
\gamma-1\right)  \frac{\partial%
\mathcal{L}%
_{j}\left(  n_{j0},T_{j0}\right)  }{T_{j0}\partial n_{j0}},\tag{A12}\\
\Omega_{ie}  & =\nu_{ie}^{\varepsilon}\left(  n_{e0},T_{e0}\right)
,\Omega_{ei}=\nu_{ei}^{\varepsilon}\left(  n_{i0},T_{e0}\right)  .\nonumber
\end{align}
From Equations (A9) and (A10), we find equations for $T_{i1}$ and $T_{e1}$%
\begin{equation}
DT_{i1}=G_{4}\mathbf{\nabla}\cdot\mathbf{v}_{i1}+G_{3}\mathbf{\nabla}%
\cdot\mathbf{v}_{e1}\tag{A13}%
\end{equation}
and%
\begin{equation}
DT_{e1}=G_{1}\mathbf{\nabla}\cdot\mathbf{v}_{e1}+G_{2}\mathbf{\nabla}%
\cdot\mathbf{v}_{i1}.\tag{A14}%
\end{equation}
Here, we have%
\begin{align}
D  & =D_{1i}D_{1e}-D_{2i}D_{2e},\tag{A15}\\
G_{1}  & =D_{1i}C_{1e},G_{2}=D_{2e}C_{1i},\nonumber\\
G_{3}  & =D_{2i}C_{1e},G_{4}=D_{1e}C_{1i}.\nonumber
\end{align}

\bigskip

\subsection{Expressions for $\mathbf{\nabla\cdot v}_{i,e1}$}

We now substitute temperature perturbations $T_{i,e1}$ defined by Equations
(A13) and (A14) into Equations (A3) and (A4). Then applying operators
$\partial/\partial x$ and $\partial/\partial y$ to Equations (A3) and (A4),
respectively, and adding them, we find equation for $\mathbf{\nabla}%
\cdot\mathbf{v}_{i1}$%
\begin{equation}
L_{1i}\mathbf{\nabla}\cdot\mathbf{v}_{i1}=-L_{2i}\mathbf{\nabla}%
\cdot\mathbf{v}_{e1}+\Phi_{i1},\tag{A16}%
\end{equation}
where
\begin{align}
L_{1i}  & =\Omega_{i}^{2}+\frac{1}{m_{i}}\left[  \frac{G_{4}}{D}\left(
\frac{\partial}{\partial t}+v_{i0y}\frac{\partial}{\partial y}\right)
-T_{i0}\right]  \mathbf{\nabla}^{2},\tag{A17}\\
L_{2i}  & =\frac{1}{m_{i}}\frac{G_{3}}{D}\left(  \frac{\partial}{\partial
t}+v_{i0y}\frac{\partial}{\partial y}\right)  \mathbf{\nabla}^{2},\nonumber\\
\Phi_{i1}  & =\omega_{ci}\left(  \frac{\partial F_{i1y}}{\partial x}%
-\frac{\partial F_{i1x}}{\partial y}\right)  +\left(  \frac{\partial}{\partial
t}+v_{i0y}\frac{\partial}{\partial y}\right)  \mathbf{\nabla\cdot F}%
_{i1}.\nonumber
\end{align}
Analogously, using Equations (A6) and (A7), we obtain
\begin{equation}
L_{1e}\mathbf{\nabla}\cdot\mathbf{v}_{e1}=-L_{2e}\mathbf{\nabla}%
\cdot\mathbf{v}_{i1}+\Phi_{e1},\tag{A18}%
\end{equation}
where%
\begin{align}
L_{1e}  & =\Omega_{e}^{2}+\frac{1}{m_{e}}\left(  \frac{G_{1}}{D}\frac
{\partial}{\partial t}-T_{e0}\right)  \mathbf{\nabla}^{2},\tag{A19}\\
L_{2e}  & =\frac{1}{m_{e}}\frac{G_{2}}{D}\frac{\partial}{\partial
t}\mathbf{\nabla}^{2},\nonumber\\
\Phi_{e1}  & =\omega_{ce}\left(  \frac{\partial F_{e1y}}{\partial x}%
-\frac{\partial F_{e1x}}{\partial y}\right)  +\frac{\partial}{\partial
t}\mathbf{\nabla\cdot F}_{e1}.\nonumber
\end{align}
From Equations (A16) and (A18), we find
\begin{equation}
L\mathbf{\nabla}\cdot\mathbf{v}_{i1}=L_{1e}\Phi_{i1}-L_{2i}\Phi_{e1}\tag{A20}%
\end{equation}
and%
\begin{equation}
L\mathbf{\nabla}\cdot\mathbf{v}_{e1}=L_{1i}\Phi_{e1}-L_{2e}\Phi_{i1}.\tag{A21}%
\end{equation}
The operator $L$ is given by%
\begin{equation}
L=L_{1i}L_{1e}-L_{2i}L_{2e}.\tag{A22}%
\end{equation}

\bigskip

\subsection{Equations for ion and electron velocities via\textit{\ }%
$\mathbf{F}_{i,e1}$}

Using Equations (A3), (A4), (A13), (A20), and (A21), we obtain the following
equations for components of the perturbed ion velocity:%
\begin{equation}
\Omega_{i}^{2}v_{i1x}=\frac{L_{ix}}{m_{i}DL}\left(  A_{1i}\Phi_{i1}-A_{2i}%
\Phi_{e1}\right)  +\omega_{ci}F_{i1y}+\left(  \frac{\partial}{\partial
t}+v_{i0y}\frac{\partial}{\partial y}\right)  F_{i1x}\tag{A23}%
\end{equation}
and%
\begin{equation}
\Omega_{i}^{2}v_{i1y}=\frac{L_{iy}}{m_{i}DL}\left(  A_{1i}\Phi_{i1}-A_{2i}%
\Phi_{e1}\right)  -\omega_{ci}F_{i1x}+\left(  \frac{\partial}{\partial
t}+v_{i0y}\frac{\partial}{\partial y}\right)  F_{i1y}.\tag{A24}%
\end{equation}
The operators $A_{1,2i}$ are given by%
\begin{align}
A_{1i}  & =\left[  G_{4}-DT_{i0}\left(  \frac{\partial}{\partial t}%
+v_{i0y}\frac{\partial}{\partial y}\right)  ^{-1}\right]  L_{1e}-G_{3}%
L_{2e},\tag{A25}\\
A_{2i}  & =\left[  G_{4}-DT_{i0}\left(  \frac{\partial}{\partial t}%
+v_{i0y}\frac{\partial}{\partial y}\right)  ^{-1}\right]  L_{2i}-G_{3}%
L_{1i}.\nonumber
\end{align}

Equations for components of the perturbed electron velocity are found by using
Equations (A6), (A7), (A14), (A20), and (A21)%
\begin{equation}
\Omega_{e}^{2}v_{e1x}=\frac{L_{ex}}{m_{e}DL}\left(  A_{1e}\Phi_{e1}-A_{2e}%
\Phi_{i1}\right)  +\omega_{ce}F_{e1y}+\frac{\partial F_{e1x}}{\partial
t},\tag{A26}%
\end{equation}%
\begin{equation}
\Omega_{e}^{2}v_{e1y}=\frac{L_{ey}}{m_{e}DL}\left(  A_{1e}\Phi_{e1}-A_{2e}%
\Phi_{i1}\right)  -\omega_{ce}F_{e1x}+\frac{\partial F_{e1y}}{\partial
t}.\tag{A27}%
\end{equation}
Here,%
\begin{align}
A_{1e}  & =\left[  G_{1}-DT_{e0}\left(  \frac{\partial}{\partial t}\right)
^{-1}\right]  L_{1i}-G_{2}L_{2i},\tag{A28}\\
A_{2e}  & =\left[  G_{1}-DT_{e0}\left(  \frac{\partial}{\partial t}\right)
^{-1}\right]  L_{2e}-G_{2}L_{1e}.\nonumber
\end{align}

\bigskip

\subsection{Expressions for $D$ and $G_{1,2,3,4}$}

We now give expressions for $D$ and $G_{1,2,3,4}$ defined by Equation (A15).
Using Equation (A11), we find%
\begin{align}
\left(  \frac{\partial}{\partial t}+v_{i0y}\frac{\partial}{\partial y}\right)
^{-1}\left(  \frac{\partial}{\partial t}\right)  ^{-1}D  & =\left[  \left(
\frac{\partial}{\partial t}+v_{i0y}\frac{\partial}{\partial y}\right)
+\Omega_{Ti}\right]  \left(  \frac{\partial}{\partial t}+\Omega_{Te}\right)
\tag{A29}\\
& +\left(  \frac{\partial}{\partial t}+\Omega_{Te}\right)  \Omega_{ie}+\left[
\left(  \frac{\partial}{\partial t}+v_{i0y}\frac{\partial}{\partial y}\right)
+\Omega_{Ti}\right]  \Omega_{ei}\nonumber
\end{align}
and%
\begin{align}
G_{1}  & =T_{e0}\left[  \left(  \frac{\partial}{\partial t}+v_{i0y}%
\frac{\partial}{\partial y}\right)  +\Omega_{Ti}+\Omega_{ie}\right]  \left[
-\left(  \gamma-1\right)  \frac{\partial}{\partial t}+\Omega_{ne}\right]
\left(  \frac{\partial}{\partial t}+v_{i0y}\frac{\partial}{\partial y}\right)
,\tag{A30}\\
G_{2}  & =T_{i0}\Omega_{ei}\left[  -\left(  \gamma-1\right)  \left(
\frac{\partial}{\partial t}+v_{i0y}\frac{\partial}{\partial y}\right)
+\Omega_{ni}\right]  \frac{\partial}{\partial t},\nonumber\\
G_{3}  & =T_{e0}\Omega_{ie}\left[  -\left(  \gamma-1\right)  \frac{\partial
}{\partial t}+\Omega_{ne}\right]  \left(  \frac{\partial}{\partial t}%
+v_{i0y}\frac{\partial}{\partial y}\right)  ,\nonumber\\
G_{4}  & =T_{i0}\left(  \frac{\partial}{\partial t}+\Omega_{Te}+\Omega
_{ei}\right)  \left[  -\left(  \gamma-1\right)  \left(  \frac{\partial
}{\partial t}+v_{i0y}\frac{\partial}{\partial y}\right)  +\Omega_{ni}\right]
\frac{\partial}{\partial t}.\nonumber
\end{align}

\bigskip

\subsection{Simplified expressions for $A_{1,2i}$ and $A_{1,2e}$}

We can further simplify expressions for $A_{1,2i}$ and $A_{1,2e}$ given by
Equations (A25) and (A28). Using Equation (A17), we obtain%
\begin{equation}
A_{2i}=-G_{3}\Omega_{i}^{2}.\tag{A31}%
\end{equation}
The expression for $A_{1i}$ can be given in the form%
\begin{equation}
A_{1i}=\left[  G_{4}-DT_{i0}\left(  \frac{\partial}{\partial t}+v_{i0y}%
\frac{\partial}{\partial y}\right)  ^{-1}\right]  \Omega_{e}^{2}-\frac
{1}{m_{e}}\mathbf{\nabla}^{2}K,\tag{A32}%
\end{equation}
where we have used Equation (A19). The following notation is introduced in
Equation (A32):%
\begin{equation}
K=\frac{1}{D}\left(  G_{2}G_{3}-G_{1}G_{4}\right)  \frac{\partial}{\partial
t}+G_{4}T_{e0}+G_{1}T_{i0}\left(  \frac{\partial}{\partial t}+v_{i0y}%
\frac{\partial}{\partial y}\right)  ^{-1}\frac{\partial}{\partial t}%
-DT_{i0}T_{e0}\left(  \frac{\partial}{\partial t}+v_{i0y}\frac{\partial
}{\partial y}\right)  ^{-1}.\tag{A33}%
\end{equation}
Analogously, we will have%
\begin{equation}
A_{2e}=-G_{2}\Omega_{e}^{2}\tag{A34}%
\end{equation}
and%
\begin{equation}
A_{1e}=\left[  G_{1}-DT_{e0}\left(  \frac{\partial}{\partial t}\right)
^{-1}\right]  \Omega_{i}^{2}-\frac{1}{m_{i}}\mathbf{\nabla}^{2}\left(
\frac{\partial}{\partial t}+v_{i0y}\frac{\partial}{\partial y}\right)  \left(
\frac{\partial}{\partial t}\right)  ^{-1}K.\tag{A35}%
\end{equation}

Calculations show that the value $D^{-1}\left(  G_{2}G_{3}-G_{1}G_{4}\right)
$ takes the simple form%
\begin{equation}
\frac{1}{D}\left(  G_{2}G_{3}-G_{1}G_{4}\right)  =-T_{i0}T_{e0}\left[
-\left(  \gamma-1\right)  \left(  \frac{\partial}{\partial t}+v_{i0y}%
\frac{\partial}{\partial y}\right)  +\Omega_{ni}\right]  \left[  -\left(
\gamma-1\right)  \frac{\partial}{\partial t}+\Omega_{ne}\right]  .\tag{A36}%
\end{equation}
Using Equations (A29), (A30), and (A36), we can also write the value $K$
defined by Equation (A33) in the simple form%
\begin{equation}
K=-T_{i0}T_{e0}\left(  W_{i}W_{e}+W_{i}\Omega_{ei}+W_{e}\Omega_{ie}\right)
\frac{\partial}{\partial t}.\tag{A37}%
\end{equation}
Here, notations are introduced
\begin{align}
W_{i}  & =\gamma\left(  \frac{\partial}{\partial t}+v_{i0y}\frac{\partial
}{\partial y}\right)  +\Omega_{Ti}-\Omega_{ni},\tag{A38}\\
W_{e}  & =\gamma\frac{\partial}{\partial t}+\Omega_{Te}-\Omega_{ne}.\nonumber
\end{align}
We remind the reader that the temperatures of the ions and electrons are
considered to be equal one another. We retain different notations for the
control of the symmetry of the ion and electron contribution. Analogously, we
find the following values:%
\begin{align}
G_{4}-DT_{i0}\left(  \frac{\partial}{\partial t}+v_{i0y}\frac{\partial
}{\partial y}\right)  ^{-1}  & =-T_{i0}\left(  W_{i}V_{e}+W_{i}\Omega
_{ei}+V_{e}\Omega_{ie}\right)  \frac{\partial}{\partial t},\tag{A39}\\
G_{1}-DT_{e0}\left(  \frac{\partial}{\partial t}\right)  ^{-1}  &
=-T_{e0}\left(  W_{e}V_{i}+W_{e}\Omega_{ie}+V_{i}\Omega_{ei}\right)  \left(
\frac{\partial}{\partial t}+v_{i0y}\frac{\partial}{\partial y}\right)
,\nonumber
\end{align}
where%
\begin{align}
V_{i}  & =\left(  \frac{\partial}{\partial t}+v_{i0y}\frac{\partial}{\partial
y}\right)  +\Omega_{Ti},\tag{A40}\\
V_{e}  & =\frac{\partial}{\partial t}+\Omega_{Te}.\nonumber
\end{align}

\bigskip

\subsection{Operator $L$}

Let us find the operator $L$ given by Equation (A22). Using Equations (A17)
and (A19), we obtain
\begin{align}
L  & =\Omega_{i}^{2}\Omega_{e}^{2}+\frac{1}{m_{i}}\Omega_{e}^{2}\left[
\frac{G_{4}}{D}\left(  \frac{\partial}{\partial t}+v_{i0y}\frac{\partial
}{\partial y}\right)  -T_{i0}\right]  \mathbf{\nabla}^{2}+\frac{1}{m_{e}%
}\Omega_{i}^{2}\left(  \frac{G_{1}}{D}\frac{\partial}{\partial t}%
-T_{e0}\right)  \mathbf{\nabla}^{2}\tag{A41}\\
& -\frac{1}{m_{i}m_{e}D}\left(  \frac{\partial}{\partial t}+v_{i0y}%
\frac{\partial}{\partial y}\right)  \mathbf{\nabla}^{4}K.\nonumber
\end{align}
The expressions containing in this equation are given by Equations (A37)-(A40).

\bigskip

\subsection{Simplified equations for ion and electron velocities
via\textit{\ }$\mathbf{E}_{1}$}

We now substitute expressions for $A_{1,2i}$ given by Equations (A31) and
(A32) into Equations (A23) and (A24). Then, we replace the values
$\mathbf{F}_{j1}$ and $\Phi_{i,e1}$ by their expressions through
$\mathbf{E}_{1}$ which are given by
\begin{align}
F_{j1x}  & =\frac{q_{j}}{m_{j}}\left[  E_{1x}+v_{j0y}\left(  \frac{\partial
}{\partial t}\right)  ^{-1}\left(  \frac{\partial E_{1x}}{\partial y}%
-\frac{\partial E_{1y}}{\partial x}\right)  \right]  ,\tag{A42}\\
F_{j1y}  & =\frac{q_{j}}{m_{j}}E_{1y}\nonumber
\end{align}
and%
\begin{align}
\Phi_{i1}  & =-\frac{q_{i}}{m_{i}}\left(  \omega_{ci}-v_{i0y}\frac{\partial
}{\partial x}\right)  \left(  \frac{\partial}{\partial t}+v_{i0y}%
\frac{\partial}{\partial y}\right)  \left(  \frac{\partial}{\partial
t}\right)  ^{-1}\left(  \frac{\partial E_{1x}}{\partial y}-\frac{\partial
E_{1y}}{\partial x}\right) \tag{A43}\\
& +\frac{q_{i}}{m_{i}}\left(  \frac{\partial}{\partial t}+v_{i0y}%
\frac{\partial}{\partial y}\right)  \mathbf{\nabla\cdot E}_{1},\nonumber\\
\Phi_{e1}  & =-\frac{q_{e}}{m_{e}}\omega_{ce}\left(  \frac{\partial E_{1x}%
}{\partial y}-\frac{\partial E_{1y}}{\partial x}\right)  +\frac{q_{e}}{m_{e}%
}\frac{\partial}{\partial t}\mathbf{\nabla\cdot E}_{1}.\nonumber
\end{align}
When obtaining Equations (A42) and (A43), we have used Equations (A2) and (8).
As a result, we will have the following equations for $v_{i1x}$ and $v_{i1y}$:%
\begin{align}
v_{i1x}  & =-\frac{q_{i}}{m_{i}}\frac{\Omega_{e}^{2}}{\Omega_{i}^{2}}%
\frac{L_{ix}}{L}\lambda_{i}\left[  a_{i}\left(  \frac{\partial E_{1y}%
}{\partial x}-\frac{\partial E_{1x}}{\partial y}\right)  +\left(
\frac{\partial}{\partial t}+v_{i0y}\frac{\partial}{\partial y}\right)
\mathbf{\nabla\cdot E}_{1}\right] \tag{A44}\\
& +\frac{q_{e}}{m_{e}}\frac{L_{ix}}{L}\mu_{i}\left[  \omega_{ce}\left(
\frac{\partial E_{1y}}{\partial x}-\frac{\partial E_{1x}}{\partial y}\right)
+\frac{\partial}{\partial t}\mathbf{\nabla\cdot E}_{1}\right] \nonumber\\
& +\frac{q_{i}}{m_{i}\Omega_{i}^{2}}\left(  \frac{\partial}{\partial
t}+v_{i0y}\frac{\partial}{\partial y}\right)  ^{2}\left(  \frac{\partial
}{\partial t}\right)  ^{-1}E_{1x}\nonumber\\
& +\frac{q_{i}}{m_{i}\Omega_{i}^{2}}\left[  \omega_{ci}-v_{i0y}\frac{\partial
}{\partial x}\left(  \frac{\partial}{\partial t}+v_{i0y}\frac{\partial
}{\partial y}\right)  \left(  \frac{\partial}{\partial t}\right)
^{-1}\right]  E_{1y}\nonumber
\end{align}
and%
\begin{align}
v_{i1y}  & =-\frac{q_{i}}{m_{i}}\frac{\Omega_{e}^{2}}{\Omega_{i}^{2}}%
\frac{L_{iy}}{L}\lambda_{i}\left[  a_{i}\left(  \frac{\partial E_{1y}%
}{\partial x}-\frac{\partial E_{1x}}{\partial y}\right)  +\left(
\frac{\partial}{\partial t}+v_{i0y}\frac{\partial}{\partial y}\right)
\mathbf{\nabla\cdot E}_{1}\right] \tag{A45}\\
& +\frac{q_{e}}{m_{e}}\frac{L_{iy}}{L}\mu_{i}\left[  \omega_{ce}\left(
\frac{\partial E_{1y}}{\partial x}-\frac{\partial E_{1x}}{\partial y}\right)
+\frac{\partial}{\partial t}\mathbf{\nabla\cdot E}_{1}\right] \nonumber\\
& -\frac{q_{i}}{m_{i}}\frac{\omega_{ci}}{\Omega_{i}^{2}}\left(  \frac
{\partial}{\partial t}+v_{i0y}\frac{\partial}{\partial y}\right)  \left(
\frac{\partial}{\partial t}\right)  ^{-1}E_{1x}\nonumber\\
& +\frac{q_{i}}{m_{i}\Omega_{i}^{2}}\left[  \omega_{ci}v_{i0y}\frac{\partial
}{\partial x}\left(  \frac{\partial}{\partial t}\right)  ^{-1}+\left(
\frac{\partial}{\partial t}+v_{i0y}\frac{\partial}{\partial y}\right)
\right]  E_{1y},\nonumber
\end{align}
where notations are
\begin{align}
\lambda_{i}  & =\frac{1}{m_{i}}\left[  T_{i0}\left(  \frac{\partial}{\partial
t}+v_{i0y}\frac{\partial}{\partial y}\right)  ^{-1}-\frac{G_{4}}{D}\right]
+\frac{1}{m_{e}m_{i}D\Omega_{e}^{2}}\mathbf{\nabla}^{2}K,\mu_{i}=\frac{G_{3}%
}{m_{i}D},\tag{A46}\\
a_{i}  & =\left(  \omega_{ci}-v_{i0y}\frac{\partial}{\partial x}\right)
\left(  \frac{\partial}{\partial t}+v_{i0y}\frac{\partial}{\partial y}\right)
\left(  \frac{\partial}{\partial t}\right)  ^{-1}.\nonumber
\end{align}
For the electron velocity, using Equations (A26), (A27), (A34), and (A35), we
obtain%
\begin{align}
v_{e1x}  & =-\frac{q_{e}}{m_{e}}\frac{\Omega_{i}^{2}}{\Omega_{e}^{2}}%
\frac{L_{ex}}{L}\lambda_{e}\left[  \omega_{ce}\left(  \frac{\partial E_{1y}%
}{\partial x}-\frac{\partial E_{1x}}{\partial y}\right)  +\frac{\partial
}{\partial t}\mathbf{\nabla\cdot E}_{1}\right] \tag{A47}\\
& +\frac{q_{i}}{m_{i}}\frac{L_{ex}}{L}\mu_{e}\left[  a_{i}\left(
\frac{\partial E_{1y}}{\partial x}-\frac{\partial E_{1x}}{\partial y}\right)
+\left(  \frac{\partial}{\partial t}+v_{i0y}\frac{\partial}{\partial
y}\right)  \mathbf{\nabla\cdot E}_{1}\right] \nonumber\\
& +\frac{q_{e}}{m_{e}}\frac{\omega_{ce}}{\Omega_{e}^{2}}E_{1y}+\frac{q_{e}%
}{m_{e}\Omega_{e}^{2}}\frac{\partial E_{1x}}{\partial t}\nonumber
\end{align}
and%
\begin{align}
v_{e1y}  & =-\frac{q_{e}}{m_{e}}\frac{\Omega_{i}^{2}}{\Omega_{e}^{2}}%
\frac{L_{ey}}{L}\lambda_{e}\left[  \omega_{ce}\left(  \frac{\partial E_{1y}%
}{\partial x}-\frac{\partial E_{1x}}{\partial y}\right)  +\frac{\partial
}{\partial t}\mathbf{\nabla\cdot E}_{1}\right] \tag{A48}\\
& +\frac{q_{i}}{m_{i}}\frac{L_{ey}}{L}\mu_{e}\left[  a_{i}\left(
\frac{\partial E_{1y}}{\partial x}-\frac{\partial E_{1x}}{\partial y}\right)
+\left(  \frac{\partial}{\partial t}+v_{i0y}\frac{\partial}{\partial
y}\right)  \mathbf{\nabla\cdot E}_{1}\right] \nonumber\\
& -\frac{q_{e}}{m_{e}}\frac{\omega_{ce}}{\Omega_{e}^{2}}E_{1x}+\frac{q_{e}%
}{m_{e}\Omega_{e}^{2}}\frac{\partial E_{1y}}{\partial t},\nonumber
\end{align}
where
\begin{align}
\lambda_{e}  & =\frac{1}{m_{e}}\left[  T_{e0}\left(  \frac{\partial}{\partial
t}\right)  ^{-1}-\frac{G_{1}}{D}\right]  +\frac{1}{m_{i}m_{e}\Omega_{i}^{2}%
D}\left(  \frac{\partial}{\partial t}+v_{i0y}\frac{\partial}{\partial
y}\right)  \left(  \frac{\partial}{\partial t}\right)  ^{-1}\mathbf{\nabla
}^{2}K,\tag{A49}\\
\mu_{e}  & =\frac{G_{2}}{m_{e}D}.\nonumber
\end{align}

\bigskip

\subsection{Perturbed plasma currents}

We now make use of obtained ion and electron velocities to find perturbed
plasma currents $j_{pl1x}=\left(  q_{i}n_{i0}v_{i1x}+q_{e}n_{e0}%
v_{e1x}\right)  $ and $j_{pl1y}=\left(  q_{i}n_{i0}v_{i1y}+q_{i}n_{i1}%
v_{i0y}+q_{e}n_{e0}v_{e1y}\right)  $ in the general form. From Equations (A44)
and (A47), we will have%
\begin{align}
4\pi\left(  \frac{\partial}{\partial t}\right)  ^{-1}j_{pl1x}  & =\alpha
_{x}\left(  \frac{\partial E_{1y}}{\partial x}-\frac{\partial E_{1x}}{\partial
y}\right)  -\beta_{x}\frac{\partial E_{1y}}{\partial x}+\delta_{x}%
\mathbf{\nabla\cdot E}_{1}\tag{A50}\\
& +\frac{\omega_{pi}^{2}}{\Omega_{i}^{2}}\left(  \frac{\partial}{\partial
t}+v_{i0y}\frac{\partial}{\partial y}\right)  ^{2}\left(  \frac{\partial
}{\partial t}\right)  ^{-2}E_{1x}+\frac{\omega_{pe}^{2}}{\Omega_{e}^{2}}%
E_{1x}\nonumber\\
& +\left(  \frac{\omega_{pi}^{2}\omega_{ci}}{\Omega_{i}^{2}}+\frac{\omega
_{pe}^{2}\omega_{ce}}{\Omega_{e}^{2}}\right)  \left(  \frac{\partial}{\partial
t}\right)  ^{-1}E_{1y}.\nonumber
\end{align}
Here,%
\begin{align}
\alpha_{x}  & =\frac{1}{L}\left[  \omega_{pi}^{2}L_{ix}\left(  \frac
{q_{e}m_{i}}{q_{i}m_{e}}\mu_{i}\omega_{ce}-\frac{\Omega_{e}^{2}}{\Omega
_{i}^{2}}\lambda_{i}a_{i}\right)  +\omega_{pe}^{2}L_{ex}\left(  \frac
{q_{i}m_{e}}{q_{e}m_{i}}\mu_{e}a_{i}-\frac{\Omega_{i}^{2}}{\Omega_{e}^{2}%
}\lambda_{e}\omega_{ce}\right)  \right]  \left(  \frac{\partial}{\partial
t}\right)  ^{-1},\tag{A51}\\
\delta_{x}  & =\omega_{pi}^{2}\frac{L_{ix}}{L}\left[  \frac{q_{e}m_{i}}%
{q_{i}m_{e}}\mu_{i}-\frac{\Omega_{e}^{2}}{\Omega_{i}^{2}}\lambda_{i}\left(
\frac{\partial}{\partial t}+v_{i0y}\frac{\partial}{\partial y}\right)  \left(
\frac{\partial}{\partial t}\right)  ^{-1}\right] \nonumber\\
& +\omega_{pe}^{2}\frac{L_{ex}}{L}\left[  \frac{q_{i}m_{e}}{q_{e}m_{i}}\mu
_{e}\left(  \frac{\partial}{\partial t}+v_{i0y}\frac{\partial}{\partial
y}\right)  \left(  \frac{\partial}{\partial t}\right)  ^{-1}-\frac{\Omega
_{i}^{2}}{\Omega_{e}^{2}}\lambda_{e}\right]  ,\nonumber\\
\beta_{x}  & =\frac{\omega_{pi}^{2}}{\Omega_{i}^{2}}v_{i0y}\left(
\frac{\partial}{\partial t}+v_{i0y}\frac{\partial}{\partial y}\right)  \left(
\frac{\partial}{\partial t}\right)  ^{-2},\nonumber
\end{align}
and $\omega_{pj}=\left(  4\pi n_{j0}q_{j}^{2}/m_{j}\right)  ^{1/2}$ is the
plasma frequency. The values $\lambda_{i,e}$, $\mu_{i,e}$, and $a_{i}$ are
given by Equations (A46) and (A49). Using Equations (2), (A44), (A45), and
(A48), we further find
\begin{align}
4\pi\left(  \frac{\partial}{\partial t}\right)  ^{-1}j_{pl1y}  & =\left(
\alpha_{y}+\eta_{1}\right)  \left(  \frac{\partial E_{1y}}{\partial x}%
-\frac{\partial E_{1x}}{\partial y}\right)  -\beta_{x}\frac{\partial E_{1x}%
}{\partial x}+\beta_{y}\frac{\partial E_{1y}}{\partial x}+\left(  \delta
_{y}+\eta_{2}\right)  \mathbf{\nabla\cdot E}_{1}\tag{A52}\\
& -\left(  \frac{\omega_{pi}^{2}\omega_{ci}}{\Omega_{i}^{2}}+\frac{\omega
_{pe}^{2}\omega_{ce}}{\Omega_{e}^{2}}\right)  \left(  \frac{\partial}{\partial
t}\right)  ^{-1}E_{1x}+\left(  \frac{\omega_{pi}^{2}}{\Omega_{i}^{2}}%
+\frac{\omega_{pe}^{2}}{\Omega_{e}^{2}}\right)  E_{1y},\nonumber
\end{align}
where%
\begin{align}
\alpha_{y}  & =\omega_{pi}^{2}\frac{L_{iy}}{L}\left(  \frac{q_{e}m_{i}}%
{q_{i}m_{e}}\mu_{i}\omega_{ce}-\frac{\Omega_{e}^{2}}{\Omega_{i}^{2}}%
\lambda_{i}a_{i}\right)  \left(  \frac{\partial}{\partial t}+v_{i0y}%
\frac{\partial}{\partial y}\right)  ^{-1}\tag{A53}\\
& +\omega_{pe}^{2}\frac{L_{ey}}{L}\left(  \frac{q_{i}m_{e}}{q_{e}m_{i}}\mu
_{e}a_{i}-\frac{\Omega_{i}^{2}}{\Omega_{e}^{2}}\lambda_{e}\omega_{ce}\right)
\left(  \frac{\partial}{\partial t}\right)  ^{-1},\nonumber\\
\delta_{y}  & =\omega_{pi}^{2}\frac{L_{iy}}{L}\left[  \frac{q_{e}m_{i}}%
{q_{i}m_{e}}\mu_{i}\left(  \frac{\partial}{\partial t}+v_{i0y}\frac{\partial
}{\partial y}\right)  ^{-1}\frac{\partial}{\partial t}-\frac{\Omega_{e}^{2}%
}{\Omega_{i}^{2}}\lambda_{i}\right] \nonumber\\
& +\omega_{pe}^{2}\frac{L_{ey}}{L}\left[  \frac{q_{i}m_{e}}{q_{e}m_{i}}\mu
_{e}\left(  \frac{\partial}{\partial t}+v_{i0y}\frac{\partial}{\partial
y}\right)  \left(  \frac{\partial}{\partial t}\right)  ^{-1}-\frac{\Omega
_{i}^{2}}{\Omega_{e}^{2}}\lambda_{e}\right]  ,\nonumber\\
\eta_{1}  & =\omega_{pi}^{2}v_{i0y}\frac{L_{ix}}{L}\left(  \frac{\Omega
_{e}^{2}}{\Omega_{i}^{2}}\lambda_{i}a_{i}-\frac{q_{e}m_{i}}{q_{i}m_{e}}\mu
_{i}\omega_{ce}\right)  \left(  \frac{\partial}{\partial t}+v_{i0y}%
\frac{\partial}{\partial y}\right)  ^{-1}\left(  \frac{\partial}{\partial
t}\right)  ^{-1}\frac{\partial}{\partial x},\nonumber\\
\eta_{2}  & =\omega_{pi}^{2}v_{i0y}\frac{L_{ix}}{L}\left[  \frac{\Omega
_{e}^{2}}{\Omega_{i}^{2}}\lambda_{i}\left(  \frac{\partial}{\partial
t}\right)  ^{-1}-\frac{q_{e}m_{i}}{q_{i}m_{e}}\mu_{i}\left(  \frac{\partial
}{\partial t}+v_{i0y}\frac{\partial}{\partial y}\right)  ^{-1}\right]
\frac{\partial}{\partial x},\nonumber\\
\beta_{y}  & =\frac{\omega_{pi}^{2}}{\Omega_{i}^{2}}v_{i0y}^{2}\left(
\frac{\partial}{\partial t}\right)  ^{-2}\frac{\partial}{\partial x}.\nonumber
\end{align}

We can rewrite Equations (A50) and (A52) in the form%
\begin{equation}
4\pi\left(  \frac{\partial}{\partial t}\right)  ^{-1}j_{pl1x}=\varepsilon
_{plxx}E_{1x}+\varepsilon_{plxy}E_{1y}\tag{A54}%
\end{equation}
and%
\begin{equation}
4\pi\left(  \frac{\partial}{\partial t}\right)  ^{-1}j_{pl1y}=\varepsilon
_{plyx}E_{1x}+\varepsilon_{plyy}E_{1y},\tag{A55}%
\end{equation}
where the components of the plasma dielectric permeability tensor are given by%
\begin{align}
\varepsilon_{plxx}  & =-\alpha_{x}\frac{\partial}{\partial y}+\delta_{x}%
\frac{\partial\ }{\partial x}+\frac{\omega_{pi}^{2}}{\Omega_{i}^{2}}\left(
\frac{\partial}{\partial t}+v_{i0y}\frac{\partial}{\partial y}\right)
^{2}\left(  \frac{\partial}{\partial t}\right)  ^{-2}+\frac{\omega_{pe}^{2}%
}{\Omega_{e}^{2}},\tag{A56}\\
\varepsilon_{plxy}  & =\alpha_{x}\frac{\partial}{\partial x}-\beta_{x}%
\frac{\partial}{\partial x}+\delta_{x}\frac{\partial}{\partial y}+\left(
\frac{\omega_{pi}^{2}\omega_{ci}}{\Omega_{i}^{2}}+\frac{\omega_{pe}^{2}%
\omega_{ce}}{\Omega_{e}^{2}}\right)  \left(  \frac{\partial}{\partial
t}\right)  ^{-1},\nonumber\\
\varepsilon_{plyx}  & =-\left(  \alpha_{y}+\eta_{1}\right)  \frac{\partial
}{\partial y}-\beta_{x}\frac{\partial}{\partial x}+\left(  \delta_{y}+\eta
_{2}\right)  \frac{\partial}{\partial x}-\left(  \frac{\omega_{pi}^{2}%
\omega_{ci}}{\Omega_{i}^{2}}+\frac{\omega_{pe}^{2}\omega_{ce}}{\Omega_{e}^{2}%
}\right)  \left(  \frac{\partial}{\partial t}\right)  ^{-1},\nonumber\\
\varepsilon_{plyy}  & =\left(  \alpha_{y}+\eta_{1}\right)  \frac{\partial
}{\partial x}+\beta_{y}\frac{\partial}{\partial x}+\left(  \delta_{y}+\eta
_{2}\right)  \frac{\partial}{\partial y}+\frac{\omega_{pi}^{2}}{\Omega_{i}%
^{2}}+\frac{\omega_{pe}^{2}}{\Omega_{e}^{2}}.\nonumber
\end{align}
Using Equations (A51) and (A53), we can find $\varepsilon_{plij}$ in specific cases.

\bigskip

\section{Appendix}

\subsection{Perturbed velocity of cosmic rays}

The linearized Equation (5) for the cold, nonrelativistic, $T_{cr}\ll
m_{cr}c^{2}$, cosmic rays takes the form%
\begin{equation}
\gamma_{cr0}\left(  \frac{\partial}{\partial t}+u_{cr}\frac{\partial}{\partial
y}\right)  \left(  \mathbf{v}_{cr1}+\gamma_{cr0}^{2}\frac{\mathbf{u}%
_{cr}u_{cr}}{c^{2}}v_{cr1y}\right)  =-\frac{\mathbf{\nabla}p_{cr1}}%
{m_{cr}n_{cr0}}+\mathbf{F}_{cr1}+\frac{q_{cr}}{m_{cr}c}\mathbf{v}_{cr1}%
\times\mathbf{B}_{0},\tag{B1}%
\end{equation}
where%
\begin{equation}
\mathbf{F}_{cr1}=\frac{q_{cr}}{m_{cr}}\left(  \mathbf{E}_{1}\mathbf{+}\frac
{1}{c}\mathbf{u}_{cr}\times\mathbf{B}_{1}\right)  .\tag{B2}%
\end{equation}
When obtaining Equation (B1), we have used that $\mathbf{u}_{cr}$ is directed
along the $y$-axis and $\gamma_{cr1}=\gamma_{cr0}^{3}u_{cr}v_{cr1y}/c^{2}$,
where $\gamma_{cr0}=\left(  1-u_{cr}^{2}/c^{2}\right)  ^{-1/2}$. From Equation
(B1), we find the following equations for $v_{cr1x,y}$:%
\begin{equation}
\gamma_{cr0}\left(  \frac{\partial}{\partial t}+u_{cr}\frac{\partial}{\partial
y}\right)  v_{cr1x}=-\frac{1}{m_{cr}n_{cr0}}\frac{\partial p_{cr1}}{\partial
x}+F_{cr1x}+\omega_{ccr}v_{cr1y}\ \tag{B3}%
\end{equation}
and
\begin{equation}
\gamma_{cr0}^{3}\left(  \frac{\partial}{\partial t}+u_{cr}\frac{\partial
}{\partial y}\right)  v_{cr1y}=-\frac{1}{m_{cr}n_{cr0}}\frac{\partial p_{cr1}%
}{\partial y}+F_{cr1y}-\omega_{ccr}v_{cr1x},\tag{B4}%
\end{equation}
where $\omega_{ccr}=q_{cr}B_{0}/m_{cr}c$ is the cyclotron frequency of the
cosmic ray particles. Solutions of Equations (B3) and (B4) have the form%
\begin{equation}
\Omega_{cr}^{2}v_{cr1x}=\frac{1}{m_{cr}n_{cr0}}L_{1crx}p_{cr1}+\omega
_{ccr}\ F_{cr1y}+\gamma_{cr0}^{3}\left(  \frac{\partial}{\partial t}%
+u_{cr}\frac{\partial}{\partial y}\right)  F_{cr1x}\tag{B5}%
\end{equation}
and%
\begin{equation}
\Omega_{cr}^{2}v_{cr1y}=\frac{1}{m_{cr}n_{cr0}}L_{1cry}p_{cr1}-\omega
_{ccr}F_{cr1x}+\gamma_{cr0}\left(  \frac{\partial}{\partial t}+u_{cr}%
\frac{\partial}{\partial y}\right)  F_{cr1y},\tag{B6}%
\end{equation}
where
\begin{align}
\Omega_{cr}^{2}  & =\gamma_{cr0}^{4}\left(  \frac{\partial}{\partial t}%
+u_{cr}\frac{\partial}{\partial y}\right)  ^{2}+\omega_{ccr}^{2},\tag{B7}\\
L_{1crx}  & =-\omega_{ccr}\ \frac{\partial}{\partial y}-\gamma_{cr0}%
^{3}\left(  \frac{\partial}{\partial t}+u_{cr}\frac{\partial}{\partial
y}\right)  \frac{\partial}{\partial x},\nonumber\\
L_{1cry}  & =\omega_{ccr}\frac{\partial}{\partial x}-\gamma_{cr0}\left(
\frac{\partial}{\partial t}+u_{cr}\frac{\partial}{\partial y}\right)
\frac{\partial}{\partial y}.\nonumber
\end{align}

\bigskip

\subsection{Equation for perturbed cosmic ray pressure}

From Equation (6) in the linear approximation, we obtain the perturbed cosmic
ray pressure%
\begin{equation}
p_{cr1}=p_{cr0}\Gamma_{cr}\left(  \frac{n_{cr1}}{n_{cr0}}-\frac{\gamma_{cr1}%
}{\gamma_{cr0}}\right)  .\tag{B8}%
\end{equation}
Using the linearized continuity equation (2) for cosmic rays and expression
for $\gamma_{cr1}$, we find that $p_{cr1}$ is given by
\begin{equation}
p_{cr1}=-p_{cr0}\Gamma_{cr}\left[  \left(  \frac{\partial}{\partial t}%
+u_{cr}\frac{\partial}{\partial y}\right)  ^{-1}\mathbf{\nabla\cdot v}%
_{cr1}+\gamma_{cr0}^{2}\frac{u_{cr}}{c^{2}}v_{cr1y}\right]  .\tag{B9}%
\end{equation}
From Equations (B5) and (B6), we obtain the expression for $\mathbf{\nabla
\cdot v}_{cr1}$ which is substituted together with the velocity $v_{cr1y}$
into Equation (B9). As a result, we have%
\begin{equation}
L_{2cr}p_{cr1}=-p_{cr0}\Gamma_{cr}\Phi_{cr1}.\tag{B10}%
\end{equation}
Here,%
\begin{align}
L_{2cr}  & =\Omega_{cr}^{2}-\gamma_{cr0}c_{scr}^{2}L_{1cr}+\gamma_{cr0}%
^{2}\frac{u_{cr}}{c^{2}}c_{scr}^{2}L_{1cry},\tag{B11}\\
\Phi_{cr1}  & =-L_{3crx}F_{cr1x}+L_{3cry}F_{cr1y},\nonumber
\end{align}
where $c_{scr}=\left(  p_{cr0}\Gamma_{cr}/m_{cr}n_{cr0}\right)  ^{1/2}$ is the
cosmic ray sound speed defined by the rest mass and%
\begin{align}
L_{1cr}  & =\gamma_{cr0}^{2}\frac{\partial^{2}}{\partial x^{2}}+\frac
{\partial^{2}}{\partial y^{2}},\tag{B12}\\
L_{3crx}  & =\omega_{ccr}\gamma_{cr0}^{2}\left(  \frac{u_{cr}}{c^{2}}%
\frac{\partial}{\partial t}+\frac{\partial}{\partial y}\right)  \left(
\frac{\partial}{\partial t}+u_{cr}\frac{\partial}{\partial y}\right)
^{-1}-\gamma_{cr0}^{3}\frac{\partial}{\partial x},\nonumber\\
L_{3cry}  & =\omega_{ccr}\left(  \frac{\partial}{\partial t}+u_{cr}%
\frac{\partial}{\partial y}\right)  ^{-1}\ \frac{\partial}{\partial x}%
+\gamma_{cr0}^{3}\left(  \frac{u_{cr}}{c^{2}}\frac{\partial}{\partial t}%
+\frac{\partial}{\partial y}\right)  .\nonumber
\end{align}

\bigskip

\subsection{Equations for cosmic ray velocities via\textit{\ }%
$\mathbf{F}_{cr1}$}

Substituting Equations (B10) and (B11) into Equations (B5) and (B6), we find%
\begin{equation}
\Omega_{cr}^{2}v_{cr1x}=\left[  c_{scr}^{2}\frac{L_{1crx}}{L_{2cr}}%
L_{3crx}+\gamma_{cr0}^{3}\left(  \frac{\partial}{\partial t}+u_{cr}%
\frac{\partial}{\partial y}\right)  \right]  F_{cr1x}+\left(  -c_{scr}%
^{2}\frac{L_{1crx}}{L_{2cr}}L_{3cry}+\omega_{ccr}\right)  \ F_{cr1y}\tag{B13}%
\end{equation}
and%
\begin{equation}
\Omega_{cr}^{2}v_{cr1y}=\left(  c_{scr}^{2}\frac{L_{1cry}}{L_{2cr}}%
L_{3crx}-\omega_{ccr}\right)  F_{cr1x}+\left[  -c_{scr}^{2}\frac{L_{1cry}%
}{L_{2cr}}L_{3cry}+\gamma_{cr0}\left(  \frac{\partial}{\partial t}+u_{cr}%
\frac{\partial}{\partial y}\right)  \right]  F_{cr1y}.\tag{B14}%
\end{equation}

\bigskip

\subsection{Equations for cosmic ray velocities via\textit{\ }%
$\mathbf{E}_{1}$}

From Equation (B2), we have%
\begin{align}
F_{cr1x}  & =\frac{q_{cr}}{m_{cr}}\left[  E_{1x}+u_{cr}\left(  \frac{\partial
}{\partial t}\right)  ^{-1}\left(  \frac{\partial E_{1x}}{\partial y}%
-\frac{\partial E_{1y}}{\partial x}\right)  \right]  ,\tag{B15}\\
F_{cr1y}  & =\frac{q_{cr}}{m_{cr}}E_{1y}.\nonumber
\end{align}
Substituting Equation (B15) into Equations (B13) and (B14), we obtain%
\begin{align}
v_{cr1x}  & =\frac{q_{cr}}{m_{cr}\Omega_{cr}^{2}}\left[  a_{crx}+\gamma
_{cr0}^{3}\left(  \frac{\partial}{\partial t}+u_{cr}\frac{\partial}{\partial
y}\right)  ^{2}\left(  \frac{\partial}{\partial t}\right)  ^{-1}\right]
E_{1x}\tag{B16}\\
& +\frac{q_{cr}}{m_{cr}\Omega_{cr}^{2}}\left[  -b_{crx}+\omega_{ccr}%
-\gamma_{cr0}^{3}u_{cr}\frac{\partial}{\partial x}\left(  \frac{\partial
}{\partial t}+u_{cr}\frac{\partial}{\partial y}\right)  \left(  \frac
{\partial}{\partial t}\right)  ^{-1}\right]  E_{1y}\nonumber
\end{align}
and%
\begin{align}
v_{cr1y}  & =\frac{q_{cr}}{m_{cr}\Omega_{cr}^{2}}\left[  a_{cry}-\omega
_{ccr}\left(  \frac{\partial}{\partial t}+u_{cr}\frac{\partial}{\partial
y}\right)  \left(  \frac{\partial}{\partial t}\right)  ^{-1}\right]
E_{1x}\tag{B17}\\
& +\frac{q_{cr}}{m_{cr}\Omega_{cr}^{2}}\left[  -b_{cry}+\omega_{ccr}%
u_{cr}\frac{\partial}{\partial x}\left(  \frac{\partial}{\partial t}\right)
^{-1}+\gamma_{cr0}\left(  \frac{\partial}{\partial t}+u_{cr}\frac{\partial
}{\partial y}\right)  \right]  E_{1y},\nonumber
\end{align}
where%
\begin{align}
a_{crx}  & =c_{scr}^{2}\frac{L_{1crx}}{L_{2cr}}L_{3crx}\left(  \frac{\partial
}{\partial t}+u_{cr}\frac{\partial}{\partial y}\right)  \left(  \frac
{\partial}{\partial t}\right)  ^{-1},\tag{B18}\\
b_{crx}  & =c_{scr}^{2}\frac{L_{1crx}}{L_{2cr}}\left[  L_{3cry}+L_{3crx}%
u_{cr}\frac{\partial}{\partial x}\left(  \frac{\partial}{\partial t}\right)
^{-1}\right]  ,\nonumber\\
a_{cry}  & =c_{scr}^{2}\frac{L_{1cry}}{L_{2cr}}L_{3crx}\left(  \frac{\partial
}{\partial t}+u_{cr}\frac{\partial}{\partial y}\right)  \left(  \frac
{\partial}{\partial t}\right)  ^{-1},\nonumber\\
b_{cry}  & =c_{scr}^{2}\frac{L_{1cry}}{L_{2cr}}\left[  L_{3cry}+L_{3crx}%
u_{cr}\frac{\partial}{\partial x}\left(  \frac{\partial}{\partial t}\right)
^{-1}\right]  .\nonumber
\end{align}
The operators $L_{1crx,y}$, $L_{2cr}$, and $L_{3crx,y}$ containing in Equation
(B18) are given by Equations (B7), (B11), and (B12), respectively.

\bigskip

\subsection{Perturbed cosmic ray current}

We now find the components of the perturbed cosmic ray current $j_{cr1x}%
=q_{cr}n_{cr0}v_{cr1x}$ and $j_{cr1y}=$ $q_{cr}n_{cr0}v_{cr1y}+q_{cr}%
n_{cr1}u_{cr}$. Using Equations (B16) and (B17) and continuity equation (2) in
the linear approximation, we find%
\begin{equation}
4\pi\left(  \frac{\partial}{\partial t}\right)  ^{-1}j_{cr1x}=\varepsilon
_{crxx}E_{1x}+\varepsilon_{crxy}E_{1y}\tag{B19}%
\end{equation}
and
\begin{equation}
4\pi\left(  \frac{\partial}{\partial t}\right)  ^{-1}j_{cr1y}=\varepsilon
_{cryx}E_{1x}+\varepsilon_{cryy}E_{1y}.\tag{B20}%
\end{equation}
The components of the dielectric permeability tensor are the following:%
\begin{align}
\varepsilon_{crxx}  & =\frac{\omega_{pcr}^{2}}{\Omega_{cr}^{2}}\left[
a_{crx}+\gamma_{cr0}^{3}\left(  \frac{\partial}{\partial t}+u_{cr}%
\frac{\partial}{\partial y}\right)  ^{2}\left(  \frac{\partial}{\partial
t}\right)  ^{-1}\right]  \left(  \frac{\partial}{\partial t}\right)
^{-1},\tag{B21}\\
\varepsilon_{crxy}  & =\frac{\omega_{pcr}^{2}}{\Omega_{cr}^{2}}\left[
-b_{crx}+\omega_{ccr}-\gamma_{cr0}^{3}u_{cr}\frac{\partial}{\partial x}\left(
\frac{\partial}{\partial t}+u_{cr}\frac{\partial}{\partial y}\right)  \left(
\frac{\partial}{\partial t}\right)  ^{-1}\right]  \left(  \frac{\partial
}{\partial t}\right)  ^{-1},\nonumber\\
\varepsilon_{cryx}  & =\frac{\omega_{pcr}^{2}}{\Omega_{cr}^{2}}\left[  \left(
a_{cry}\frac{\partial}{\partial t}-a_{crx}u_{cr}\frac{\partial}{\partial
x}\right)  \left(  \frac{\partial}{\partial t}+u_{cr}\frac{\partial}{\partial
y}\right)  ^{-1}\right]  \left(  \frac{\partial}{\partial t}\right)
^{-1}\nonumber\\
& -\frac{\omega_{pcr}^{2}}{\Omega_{cr}^{2}}\left[  \omega_{ccr}+\gamma
_{cr0}^{3}u_{cr}\frac{\partial}{\partial x}\left(  \frac{\partial}{\partial
t}+u_{cr}\frac{\partial}{\partial y}\right)  \left(  \frac{\partial}{\partial
t}\right)  ^{-1}\right]  \left(  \frac{\partial}{\partial t}\right)
^{-1},\nonumber\\
\varepsilon_{cryy}  & =\frac{\omega_{pcr}^{2}}{\Omega_{cr}^{2}}\left[  \left(
-b_{cry}\frac{\partial}{\partial t}+b_{crx}u_{cr}\frac{\partial}{\partial
x}\right)  \left(  \frac{\partial}{\partial t}+u_{cr}\frac{\partial}{\partial
y}\right)  ^{-1}\right]  \left(  \frac{\partial}{\partial t}\right)
^{-1}\nonumber\\
& +\frac{\omega_{pcr}^{2}}{\Omega_{cr}^{2}}\gamma_{cr0}\left[  1+\gamma
_{cr0}^{2}u_{cr}^{2}\frac{\partial^{2}}{\partial x^{2}}\left(  \frac{\partial
}{\partial t}\right)  ^{-2}\right]  .\nonumber
\end{align}

\end{appendix}

\bigskip

\bigskip

\bigskip
\end{document}